\documentclass{article}

\usepackage{arxiv}

\usepackage[utf8]{inputenc} 
\usepackage[T1]{fontenc}    
\usepackage{hyperref}       
\usepackage{url}            
\usepackage{booktabs}       
\usepackage{amsfonts}       
\usepackage{nicefrac}       
\usepackage{microtype}      
\usepackage{lipsum}
\usepackage{graphicx}
\usepackage{enumitem}
\usepackage{gensymb}
\usepackage{amsmath}
\usepackage{amssymb}
\usepackage{float}

\usepackage{mathtools}

\usepackage[ruled,linesnumbered]{algorithm2e}
\makeatletter
\newcommand{\nosemic}{\renewcommand{\@endalgocfline}{\relax}}
\newcommand{\dosemic}{\renewcommand{\@endalgocfline}{\algocf@endline}}
\let\oldnl\nl
\newcommand{\nonl}{\renewcommand{\nl}{\let\nl\oldnl}}
\makeatother

\usepackage[style=numeric-comp, sorting=none]{biblatex}
\title{Particle tracking velocimetry and trajectory curvature statistics for particle-laden liquid metal flow in the wake of a cylindrical obstacle}

\author{
  Mihails Birjukovs\\
  Institute of Numerical Modelling\\
  University of Latvia (UL)\\
  Riga, Latvia, Jelgavas 3, 1004 \\
  \texttt{mihails.birjukovs@lu.lv} \\
   \And
  Peteris Zvejnieks\\
  Institute of Numerical Modelling\\
  University of Latvia (UL)\\
  Riga, Latvia, Jelgavas 3, 1004 \\
  \texttt{peteris.zvejnieks@lu.lv} \\
   \And
  Tobias Lappan\\
  Helmholtz-Zentrum Dresden-Rossendorf (HZDR)\\
  Department of Magnetohydrodynamics\\
  Department of Transport Processes at Interfaces\\
  Bautzner Landstraße 400, 01328 Dresden, Germany \\
  \And
  Martins Klevs\\
  Institute of Numerical Modelling\\
  University of Latvia (UL)\\
  Riga, Latvia, Jelgavas 3, 1004 \\
  \texttt{peteris.zvejnieks@lu.lv} \\
   \And
  Sascha Heitkam\\
  Helmholtz-Zentrum Dresden-Rossendorf (HZDR)\\
  Department of Transport Processes at Interfaces\\
  Bautzner Landstraße 400, 01328 Dresden, Germany \\
  Technische Universität Dresden \\
  Institute of Process Engineering and Environmental Technology\\
  01062 Dresden, Germany
  \And
  Pavel Trtik \\
  Laboratory for Neutron Scattering and Imaging\\
  Paul Scherrer Institut\\
  Villigen, Switzerland, Forschungsstrasse 111, 5232 \\
  \texttt{pavel.trtik@psi.ch} \\
  \And
  David Mannes \\
  Laboratory for Neutron Scattering and Imaging\\
  Paul Scherrer Institut\\
  Villigen, Switzerland, Forschungsstrasse 111, 5232 \\
  \And
    Sven Eckert\\
  Helmholtz-Zentrum Dresden-Rossendorf (HZDR)\\
  Department of Magnetohydrodynamics\\
  Bautzner Landstraße 400, 01328 Dresden, Germany \\
   \And
    Andris Jakovics\\
  Institute of Numerical Modelling\\
  University of Latvia (UL)\\
  Riga, Latvia, Jelgavas 3, 1004
  }

\begin{document}
\maketitle

\clearpage

\begin{abstract}
This paper presents the analysis of the particle-laden liquid metal wake flow around a cylindrical obstacle at different obstacle Reynolds numbers. Particles in liquid metal are visualized using dynamic neutron radiography. We present the results of particle tracking velocimetry of the obstacle wake flow using an improved version of our image processing and particle tracking code MHT-X. The latter now utilizes divergence-free interpolation of the particle image velocimetry field used for motion prediction and particle trajectory reconstruction. We demonstrate significant improvement over the previously obtained results by showing the capabilities to assess both temporal and spatial characteristics of turbulent liquid metal flow, and validating the precision and accuracy of our methods against theoretical expectations, numerical simulations and experiments reported in literature. We obtain the expected vortex shedding frequency scaling with the obstacle Reynolds number and correctly identify the universal algebraic growth laws predicted and observed for trajectory curvature in isotropic homogeneous two-dimensional turbulence. Particle residence times within the obstacle wake and velocity statistics are also derived and found to be physically sound. Finally, we outline potential improvements to our methodology and plans for further research using neutron imaging of particle-laden flow.
\end{abstract}


\section{Introduction}
\label{sec:intro}

Bubble interaction with particles is of interest in metal purification \cite{metal-strring-1, metal-strring-2, metal-strring-3, metal-strring-4, metal-strring-5} and froth flotation \cite{flotation-book-1, flotation-book-2, sommer-4d-ptv} where gas bubbles are injected to remove impurities, which occur in form of solid particles, from the melt. Purification is accomplished mainly via two mechanisms: firstly, the rising bubbles generate turbulent flow which induces particle agglomeration, increasing the effective particle size and enhancing gravitational separation due to density differences; secondly, direct bubble-particle collisions can occur, trapping the latter at the gas-liquid interface of the bubbles ascending to the free surface. Experimental investigation of such bubble-particle interactions is the primary overarching goal of the presented research. In the first step of our investigations the focus is on the agglomeration of particles in turbulent flow. The wake flow of bubbles (i.e., the flow pattern that forms behind the bubble) is what primarily determines the trajectories of ascending bubbles if their collective effects are negligible \cite{uttt-path-instability, prl-path-instability, spiral-to-zigzag-explained, spiral-to-zigzag-explained-2}. It has also been demonstrated that free-moving particles are trapped in this wake region, increasing their local concentration and the probability of collision and agglomeration \cite{particle-boundary-conditions, particles-plus-bubbles-simulations, particles-multiscale-simulations, particle-aggregation-simulations}.

Previously we showcased a model experiment designed to study the behavior of the particles in such a wake using dynamic neutron radiography \cite{lappan2020a}. To demonstrate the possibility of tracking tiny particles in turbulent liquid metal flow, the experimental setup was limited to a quasi-two-dimensional flow channel and the configuration of a flow around a stationary cylindrical obstacle was chosen. We have developed the necessary tools to detect particles in neutron images stemming from such an experiment and demonstrated detection reliability under adverse imaging conditions \cite{birjukovs-particle-EXIF}. In addition, we have extended our object tracking code MHT-X, based on the combination of offline multiple hypothesis tracking (MHT) \cite{mht-reid-og}, Algorithm X \cite{knuth} and physics-based motion constraints, to particle tracking problems, and included the option of motion prediction assisted by particle image velocimetry (PIV) \cite{mht-x-og, birjukovs-particle-EXIF}. The code showed good performance in that it was capable of reconstructing trajectories long enough so that at least local particle tracking velocimetry (PTV), i.e. explicit velocimetry, is feasible and successfully captured the main features of the frequency spectrum of particle velocity within the wake. However, we identified several aspects of the code that needed improvement so that particles could be more reliably tracked throughout the entire imaged field of view (FOV) -- namely, PIV field interpolation and trajectory interpolation were the main components to be addressed for better tracking performance. Of course, it must also be clear that the shape of larger ascending gas bubbles is not a sphere and their trajectories are not rectilinear, but rather zigzagging or helical -- this is to be accounted for in the future studies.

 Before explaining the improvements to the code and how they enable a more in-depth analysis of particle flow in liquid metal, we must first note that, despite the physical interest, there is very little experimental work (in contrast to many available simulations \cite{hzdr-ibm-bubbles-thesis, dns-longitudinal-field, imb-transverse-field, zhang-mf-vertical, zhang-mf-simulations, gaudlitz-shape-wake-variations-1-bubble, hele-shaw-bubbles-vof, hele-shaw-bubbles-experiment}) where bubble wakes or bubble/particle interactions are directly visualized in liquid metal \cite{lappan2020a, birjukovs-particle-EXIF}. One of the main reasons for this is the lack of suitable measurement techniques to perform such measurements in opaque liquids (in this case metals), where optical methods cannot be leveraged. Ultrasound Doppler velocimetry (UDV) has been applied to bubble wake flow characterization \cite{zhang-thesis, udv-wake-flow-structure}, but, despite sufficient temporal resolution, currently the spatial resolution is not enough to reliably identify individual particles. Particle tracking in liquid metal using positron emission particle tracking (PEPT) has also been considered for flow analysis \cite{sommer-pept, pept-1, pept-2, pept-3, pept-4}, but this method provides very low temporal resolution, making it not feasible for turbulent flows.

However, with the recent advent of dynamic X-ray and neutron radiography of two-phase liquid metal flow \cite{megumi-x-rays, saito-neutrons-1, saito-neutrons-2, lappan2020a, neutrons-particles-stirrer-scepanskis, neutrons-particles-stirrer-scepanskis-2, neutrons-simulations-stirrer-valters}, fundamental investigation of systems with multi-phase flow containing gas bubbles and solid particles is now possible, and attempts to approach industrially relevant flow conditions can be and have been made \cite{hzdr-ibm-bubbles-thesis, birjukovsArgonBubbleFlow2020, birjukovsPhaseBoundaryDynamics2020, birjukovs2021resolving, baakeNeutronRadiographyVisualization2017, x-ray-bubble-chain-simulate, x-ray-prime-code, x-ray-bubble-breakup, x-ray-bubble-coalescence, x-ray-validation, megumi-x-rays}. Despite this, the problem of proper neutron imaging of fast-moving gas bubbles has only just now been solved for bubble flow \textit{without} particles \cite{birjukovsArgonBubbleFlow2020, birjukovsPhaseBoundaryDynamics2020, birjukovs2021resolving} -- this is another reason why, in the previous studies, we have restricted ourselves to studying particle flow around a stationary cylindrical obstacle.

Another proposition made some time ago was that neutron radiography could be used to directly observe wake flow of bodies and particle flow within optically opaque systems \cite{neutrons-fluid-flow-visuals, neutrons-fluid-flow-visuals-2}. The first such benchmark study in the context of liquid metal flow with dispersed particles was recently done by Lappan et al. where gadolinium oxide particle flow around a cylindrical obstacle in a thin liquid metal channel was imaged dynamically with sufficient temporal resolution using high cold neutron flux \cite{lappan2020a, x-ray-neutron-experiments-book}. The imaged turbulent particle-laden flow was investigated using PIV and the wake flow velocity field was measured and visualized, which was then followed up by particle detection and tracking \cite{birjukovs-particle-EXIF}.

In this paper we present the improvements to the image processing and particle tracking tools that were previously developed \cite{mht-x-og, birjukovs-particle-EXIF}. Specifically, we have replaced the Delaunay triangulation-based cubic interpolation of the PIV field with divergence-free interpolation (DFI) which enforces the flow incompressibility constraint \cite{div_free, rbf_definition, multi_div_free, div-free-interpolation-siam-wendland, div-free-interpolation-particle-suspensions}. This is necessary for correct PIV-based motion prediction within the obstacle wake where the particles have lower velocity and very oscillatory trajectories, in cases when particles are entering or leaving the wake zone, as well as near the obstacle and channel walls. We have also made improvements to artifact detection and removal for the neutron images, since artifacts left over from image processing can interrupt trajectories or force incorrect reconstruction thereof. More importantly, we show that the above changes lead to significant enough increase in particle tracking quality which allows us to reconstruct many trajectories that span the entire imaging FOV; perform PTV and obtain frequency spectra and probability density functions (PDFs) for particle velocity; measure trajectory curvature $\kappa$ and derive a PDF and a FOV map for $\kappa$; assess particle residence within the FOV -- all for quasi stationary flow about a cylindrical obstacle for a range of the cylinder Reynolds number $Re_c$.

Measuring $\kappa$ in particular, but also torsion and curvature angle for particle trajectories and deriving statistics for these metrics is one of the ways of studying flow in general, and especially turbulent flow, since it yields information about the spatial scales of the flow and how they vary in time. PDFs for $\kappa$, torsion and curvature angle are known to exhibit algebraic decay with exponents depending on flow dimensionality, which is something also seen with turbulence energy spectra. There are quite a few instances where this information yields key insights into analyzed dynamical systems \cite{curvature-2d-decaying-turbulence, curvature-helicity-isotropic-turbulence, curvature-particle-turbulence-review, curvature-particle-paths-turbulent-flow, curvature-loops-in-turbulence-statistics, curvature-lagrangian-tracks-turbulence-og, 
curvature-statistical-properties-2d-turbulence, curvature-lagrangian-statistics-2d-forced-turbulence, curvature-statistics-thermal-counterflow, curvature-flow-topology, curvature-flow-topology-2, curvature-angle-stats-porous-bed, curvature-angle-stats-turbulence, curvature-arxiv-plasma-turbulence-2022, curvature-mhd-turbulence, curvature-angle-RBC}. In our case the flow is quasi two-dimensional, so torsion calculations are not possible, but we can verify that the observations from our tracking are adequate by looking at PDFs for $\kappa$. If our results are a sufficiently close to the cases where flow is comparable, it means that our methods could be applied to geometrical and statistical analysis of flow and turbulence in liquid metals based on particle flow imaging experiments. At the same time, we must point out that such experiment-based data and studies for \textit{liquid metal}, although sought after, to the authors' knowledge are currently non-existent in literature and one could so far rely only on simulations and theory.

For two-dimensional isotropic and homogeneous turbulence, it was found in \cite{curvature-lagrangian-statistics-2d-forced-turbulence} via numerical simulation with tracer particles that the algebraic decay of $\kappa$ PDFs in a circular bounded domain (forced turbulence) has a clear $k = -2.1$ exponent in vorticity-dominated regions (the Okubo-Weiss criterion is used for flow analysis). For an unbounded domain one has $k = -2.25$ instead, but this is less relevant to our case where flow in bound within a channel with a flat rectangular cross-section. On this note, optical 2D PTV was performed for thermal counterflow in a rectangular channel with a 1-$mm$ thick laser sheet, and it was found that the PDFs of $\kappa$ for Lagrangian trajectories within different particle size ranges show algebraic decay with an exponent of $k \sim - 2$ in addition to low-$\kappa$ PDF regions with $k \sim 0$ \cite{curvature-statistics-thermal-counterflow}. For the largest particles reported, which are the closest to the particle size range considered in our previous work \cite{lappan2020a, birjukovs-particle-EXIF}, the exponent is closer to $k=-2.1$, but within the error margin all particle size ranges exhibit a universal pattern. Another study involving numerical simulations of two-dimensional particle-laden (small inertial particles) forced turbulent flow, but with periodic boundary conditions, concludes roughly the same: $k = -2.07 \pm 0.09$ in the high-$\kappa$ interval of the PDF and $k = 0.0 \pm 0.1$ in the low-$\kappa$ interval, independently of the Stokes number $Stk$ \cite{curvature-statistical-properties-2d-turbulence}.  Note that in \cite{curvature-statistical-properties-2d-turbulence} and \cite{curvature-lagrangian-statistics-2d-forced-turbulence} it is assumed that particles do not affect the flow and do not interact with one another. Another study employing numerical simulations with periodic boundary conditions, this time for decaying turbulence, reports $k = -2.25$ algebraic decay that persists over time \cite{curvature-2d-decaying-turbulence}. An experimental study was performed wherein a quasi two-dimensional spatiotemporally chaotic laboratory flow in a bounded rectangular domain was analyzed with tracer particle motion imaged via their fluorescence -- algebraic decay with $k = -2$ was consistently found for different flow conditions and a $k = 0$ low-$\kappa$ PDF tail was shown \cite{curvature-flow-topology}. It was also observed that the $\kappa$ PDF $k = 0$ region is shifted towards smaller values as flow $Re$ increases and, although not noted explicitly, the $k = -2$ region of the PDF seems to be, conversely, shifted upwards with $Re$, although the increments seem smaller in comparison to the $k = 0$ case. The $k = -2$ high-$\kappa$ and $k = 0$ low-$\kappa$ power laws are further supported theoretically \cite{curvature-lagrangian-tracks-turbulence-og, curvature-flow-topology}.

With this in mind, we present the above mentioned flow characteristics for a range of $Re_c$, compare our results against existing data and also show how tracking performance of our approach changes with $Re_c$ for a fixed set of parameters for the utilized MHT-X tracking code.

\section{The experiment}

\begin{figure}[h]
\centering
\includegraphics[width=1\linewidth]{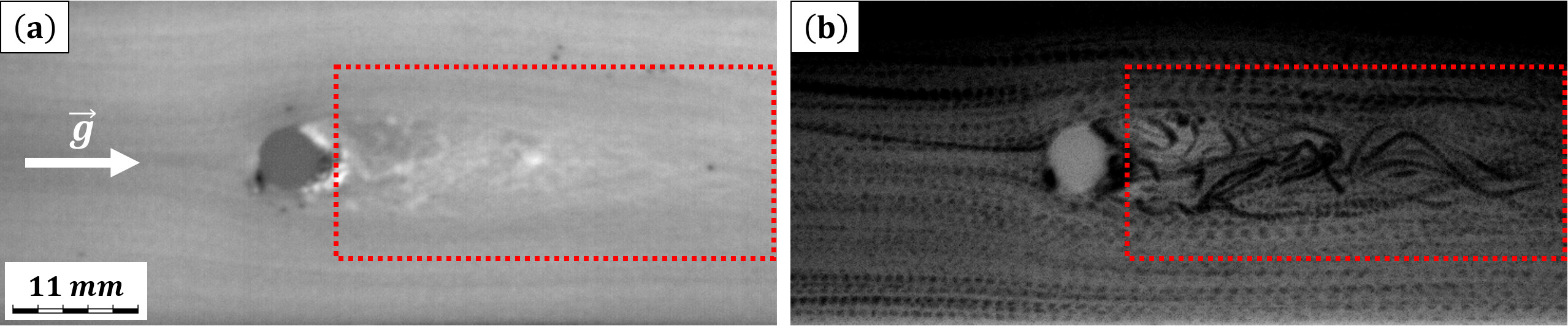}
\caption{Pixel-wise (a) standard deviation and (b) minima of luminance values within the imaged flow channel over all captured frames. The region of interest is indicated with a red dashed frame. Note the cylindrical obstacle in both figures. The images shown here were rotated 90 degrees left with respect to the originals, so here the originally downward flow is directed from left to right. The white arrow in (a) indicates the gravitational acceleration $\vec{g}$. Red dashed frames indicate the region of interest.}
\label{fig:flow-visualization-temporal-projections}
\end{figure}

A detailed description and illustration of the experimental setup used to obtain the data analyzed in this paper can be found in our previous articles \cite{lappan2020a, birjukovs-particle-EXIF}. However, for context, a brief overview is in order. Particle imaging in a low melting point gallium-tin alloy was performed using dynamic neutron radiography in a FOV centered on a straight section of a closed-loop flow channel with a uniform $30~mm \times 3~mm$ rectangular cross-section. Neutron flux was directed along the $3~mm$ dimension. The FOV contained a cylindrical obstacle (centered and fixed in the channel) with a $5~mm$ diameter (Figure \ref{fig:flow-visualization-temporal-projections}), the boundary of which obeyed the no-slip condition. Liquid metal flow was driven by a disc-type electromagnetic induction pump equipped with permanent magnets. The particles were made of gadolinium oxide and had a $d_\text{p} \in (0.3;0.5)~mm$ diameter. The utilized camera settings resulted in a $10~px/mm$ spatial resolution and a $10~ ms$ exposure time (100 frames per second) was set, sufficient to capture individual particles moving in the liquid metal flow \cite{lappan2020a}. A total of 24 image sequences of quasi stationary flow with a classic vortex street pattern (the wake region and particle tracks within can be seen in Figure \ref{fig:flow-visualization-temporal-projections}) were recorded, each with a different $Re_c$ value, resulting in a range of $Re_c \in [988;4147]$. For each image sequence, PIV was performed \cite{lappan2020a} and $Re_c$ was estimated from the PIV by averaging the velocity field in the free stream region (to the left of the cylindrical obstacle seen in Figure \ref{fig:flow-visualization-temporal-projections}) and then performing temporal averaging over the entire image sequence time interval. The frame count in an image sequence was 1500 to 3000 images. The region of interest that will be analyzed in this paper is highlighted with a red dashed frame in Figure \ref{fig:flow-visualization-temporal-projections}.

\section{Improved particle tracking}
\label{sec:placeholder}

\subsection{MHT-X}

We have previously demonstrated that the newly developed MHT-X algorithm showed promise for object trajectory reconstruction for scientific purposes, including systems where objects interact, e.g. split and merge \cite{mht-x-og, birjukovs-particle-EXIF}. Applied to tracking particles detected from neutron radiography images, this approach enabled us to resolve trajectories with satisfactory quality. However, it was also clear that there is potential for significant improvement, especially if a more in-depth analysis beyond velocimetry is desired. Here we present several major additions to MHT-X that dramatically increase tracking performance and make MHT-X even more generally applicable.

The main component of MHT-X is the offline form of the MHT object tracking approach \cite{mht-reid-og} which exhaustively finds the most likely set of associations based on predefined statistical functions that determine how well a connection complies with defined motion models. The difference between MHT and MHT-X is that the latter is feasible for larger problems because it has been optimized with Algorithm-X \cite{knuth}, which is possible for the offline form of MHT. MHT-X is applied to object detections acquired prior to tracing, which it attempts to connect in a directed graph (a \textit{trajectory graph}). Each node in the graph represents an object detection, and edges represent translations from one detection location to another. Each edge is assigned a likelihood value that signifies the confidence that the connection is true. There are also two auxiliary nodes, labeled \textit{Entry} and \textit{Exit}, which represent start and end points of all trajectories.

\subsubsection{Trajectory re-evaluation}

\begin{figure}[h]
\centering
\includegraphics[width=1\linewidth]{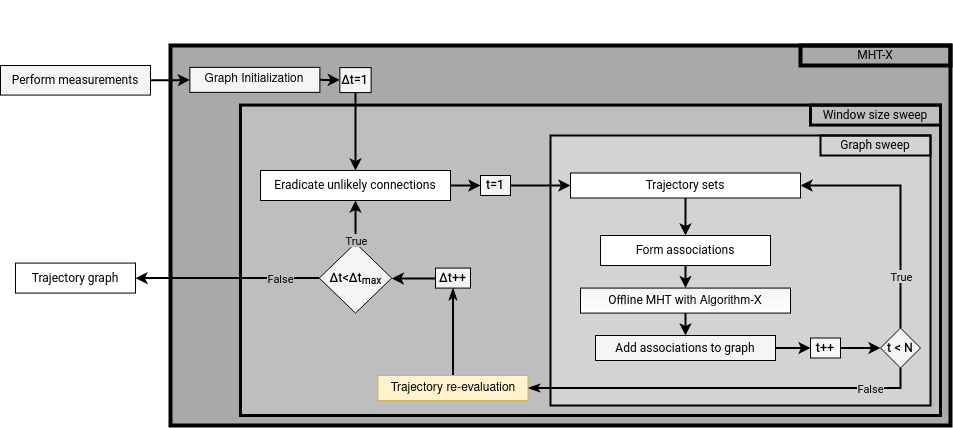}
\caption{A flowchart for the updated MHT-X algorithm (versus Figure 1 in \cite{mht-x-og}) with the new re-evaluation step.}
\label{fig:mhtx_flowchart}
\end{figure}

MHT-X performs several scans over the data set (\textit{graph sweep}) with progressively increasing time windows (\textit{window size sweep}) within which trajectory reconstruction is takes place. This enables one to employ motion models which rely on connections between objects made backwards and forwards in time, i.e. make use of \textit{context} due to the associations established between objects during preceding graph sweep iterations. Note that such information is not available \textit{a priori} from the input data. The drawback of such models is that special cases must be defined for when not enough context is available. This results in edges potentially being formed with (arguably) different models during each graph sweep. While this cannot be avoided completely, it can be remedied with trajectory re-evaluation, which is a step taken after every complete graph sweep. It recalculates the likelihood for every edge in the trajectory graph.

This step is necessary because the connections formed during earlier graph sweeps will have been formed with less context, and might, in retrospect, have overestimated likelihoods. Previously, if an absurd connection for some reason was assigned a high likelihood, it could not be eliminated because a likelihood was assigned only upon forming a connection. Now, after a trajectory is complete, it is checked how well each of its connections fits with the rest. This is done after every complete graph sweep. The procedure is as follows: every trajectory is iteratively split in two at every edge by temporarily deleting the currently tested edge. The likelihood of the temporarily deleted edge is then recalculated based on the two resulting trajectory parts by passing them to the statistical functions (statistical functions shown in \cite{mht-x-og} and \cite{birjukovs-particle-EXIF}). Similarly, the trajectory \textit{Entry} and \textit{Exit} connection likelihoods are also reevaluated.

\subsubsection{Divergence-free interpolation}

Previously, when using interpolated PIV fields for particle motion prediction, we treated the PIV vector field components as two independent scalar fields. Interpolation was necessary because PIV yielded vector values on a sparse regular grid and particle positions could be anywhere within the FOV of interest with sub-pixel coordinates. For each of the two PIV components, Delaunay triangulation was performed for the PIV point grid and cubic interpolation was used for particle centroids that were within triangles formed by nearby PIV grid points, while nearest neighbor interpolation is used otherwise. Then the interpolated velocity field was projected onto particle positions. This rather naive approach provided acceptable results.

However, here we present an improved approach using divergence-free interpolation (DFI) of the PIV data. The reason for adopting this method is that the previously used interpolation method is not physics-based and imposes no constraints on the interpolated field. Meanwhile, DFI is specifically designed such that the interpolated field analytically satisfies the flow incompressibility constraint, which was shown to have a significant impact, especially in cases with particle-laden flow \cite{div_free, rbf_definition, multi_div_free, div-free-interpolation-siam-wendland, div-free-interpolation-particle-suspensions}. In our case the flow is quasi two dimensional, so one can expect near-zero divergence.

The div-free interpolant uses matrix-valued radial basis functions (RBFs) and is described in \cite{div_free}:

\begin{equation}
\Phi(\vec{x}):=(-\nabla^2 I + \nabla\otimes \nabla)\phi_{\delta}(\vec{x})
\end{equation}
where $I$ is the identity matrix. This interpolation method is favorable because it supports arbitrarily sampled data and can be made computationally cheaper with a multilevel approach \cite{multi_div_free}. Our current non-multilevel implementation (to be upgraded to multi-level in the future) utilizes scaled RBFs:

\begin{equation}
\phi_{\delta}(\vec{x}):=\delta^{-d}\phi_{\nu, k}\left(\frac{\vert\vert\vec{x}\vert\vert}{\delta}\right)
\end{equation}
where $\delta$ is the support radius, $d$ is the number of motion dimensions of the system and $\phi_{\nu, k}$ is the RBF as defined in \cite{rbf_definition}. We used $\delta = 1100~ px$ ($102.8~mm$) and $\phi_{\nu, k}$ with $\nu = 5$ and $k = 3$ for all $Re_c$ cases.

\begin{figure}[h]
\centering
\includegraphics[width=1\linewidth]{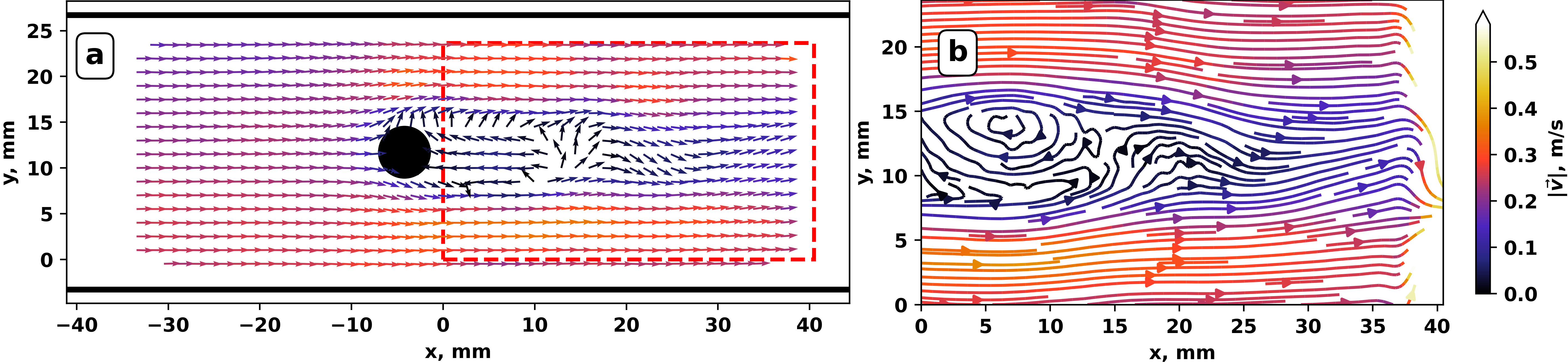}
\caption{(a) Vector field acquired with PIV. Black lines are the boundary conditions. The red striped line is the outline of the region of interest, where we performed tracing, and where the streamlines are drawn. (b) Streamlines of resulting interpolation. There is a noticeable artifact on the right side of the interpolation, where the flow abruptly changes direction and it's velocity rapidly increases (streamlines are not drawn for values too high) due to a lack of PIV data. This error is insignificant as this region is of no significance for our analysis.}
\label{fig:interpolation_example}
\end{figure}

The DFI approach also enables one to define boundary conditions (BCs) within the domain where object tracking occurs. This is important to account for because in addition to the zero-divergence constraint, the flow field must also comply with the no-slip BCs at the channel walls and the perimeter of the obstacle. The no-slip BC was set at discrete uniformly spaced points, 100 for each of the walls and 50 for the obstacle perimeter (black lines in figure \ref{fig:interpolation_example}a). The addition of this auxiliary information enables us to correctly model boundary layers near the obstacle and walls and eliminate streamline intersections with the obstacle and/or walls. That is, without the BCs for the DFI and based on MHT-X motion constraints and statistical functions alone, the particles could, for instance, "tunnel" through space where the cylindrical obstacle is.

Thus, improved tracking performance of MHT-X with DFI for the PIV field can be attributed to several factors, such as the elimination of sources and sinks in the PIV field, greater smoothness and the additional physical constraints from BCs.

\subsection{Improved artifact detection}

Previously we noted that persistent artifacts within images (e.g. stuck particles) may be a problem in that they might introduce systematic errors into trajectory reconstruction (e.g. trajectory fragmentation). We removed such artifacts with texture synthesis-based inpainting \cite{wolfram-mathematica-inpaint} after segmenting them from the inverse mean projection of the image sequence using the Otsu method \cite{otsu-thresholding}. However, as processing more data showed, this segmentation approach is not robust enough (at least for our purposes). Instead, the method is as follows (Algorithm \ref{alg:artifact-detection-method}):

\begin{algorithm}

    Compute the inverse of the mean projection for the image sequence
    
    Compute the standard deviation projection for the image sequence (Figure \ref{fig:flow-visualization-temporal-projections})
    
    Divide the output of Step 1 by that of Step 2
    
    Perform color tone mapping \cite{reproduction-of-color-chapter-6}
    
    Apply Gaussian total variation and invert the result
    
    Binarize with a manual threshold using the Otsu method output as an initial guess
    
    Apply the filling transform \cite{book-digital-image-processing} and (optionally) size thresholding
    
    Perform texture synthesis in the resulting artifact areas for all frames in an image sequence

\caption{Artifact detection}
\label{alg:artifact-detection-method}
\end{algorithm}

This, incidentally, also allows one to easily detect the obstacle in the flow channel by applying morphological opening \cite{images-mathematical-morphology} and then size thresholding to the output of Algorithm \ref{alg:artifact-detection-method}. Obstacle segmentation is necessary to prescribe the no-slip BCs for the DFI. The above method is more robust mainly because Step 3 yields a much greater contrast-to-noise ratio (CNR) for the artifacts than using just the inverse mean projection, and an optimal manual threshold can be found very quickly based on the initial Otsu method guess (which in some cases can even be enough). In certain instances, though, the CNR of the obstacle can be too low for Algorithm \ref{alg:artifact-detection-method}, so Step 6 is replaced with K-medoids clustering segmentation \cite{k-medoids-clustering}, after which one can select the level set containing the cylindrical obstacle and isolate the latter using (in this order) morphological opening, filling transform, morphological dilation \cite{images-mathematical-morphology} and again opening.

\section{Results}
\label{sec:results}

\subsection{Particle tracking performance}

To make sure the performance demonstration for the updated methods and code is fair, we decided to process all image sequences corresponding to the $Re_c \in [988;4147]$ range using identical settings for both the particle detection code and MHT-X. Image processing is done as described in \cite{birjukovs-particle-EXIF} with parameters identical to what is specified in the paper. For MHT-X, the motion models are as described in \cite{mht-x-og, birjukovs-particle-EXIF} and their parameters are as follows:

\begin{itemize}

    \item Equations (11, 12, 21) and Section 2.5.1 in \cite{mht-x-og}:
    $a_c = 12~ px/ \text{frame}^2$ (11); 
    $\lambda = 200~ px/\text{frame} $ (12); 
    $a=0.01~ px^{-1}$ and $b= width - 7\ px$: width, different for each image sequence (21); 
    $q=0.5$ (Section 2.5.1)
  
    \item Equations (7-12) in \cite{birjukovs-particle-EXIF}: 
    $\alpha = 0.3$ (7, 8); 
    $R_\text{max}=8\ px$ (8); 
    $\lambda_\text{SOI}=50\ px/\text{frame}$ (8); 
    $\sigma_\text{pos}=8\ px$ (9); 
    $\sigma_a = 15\ px/\text{frame}^2$ (10); 
    $\lambda=50\ px/\text{frame}$ (11); 
    $\beta_1=0.7$ (12); 
    $\beta_2=0.2 (12)$
  
\end{itemize}

We find that the mean particle size seen after image processing is $\langle d_\text{p} \rangle = 0.43 \pm 0.13 ~ mm$ versus the expected $d_\text{p} \in (0.3;0.5)mm$, which is in good agreement. There is a slight bias towards larger particle sizes can be explained by a lower bound on the particle size resolution imposed by the image filter settings used in and adopted from \cite{birjukovs-particle-EXIF}. Given that identical parameters are used in all cases, tracking performance degradation expected as $Re_c$ increases (particle displacements become greater, trajectory fragment connection ambiguities are more frequent) should be clearly observable. To measure the performance, we define a very strict criterion -- only trajectories with 20+ graph nodes (instances over time) are considered \textit{eligible trajectories}. It is strict in that such trajectories will be either the ones that travel past the wake flow zone or interact with it without particle entrapment, and extend throughout the entire FOV length (expected with the chosen frame rate and characteristic particle velocity \cite{birjukovs-particle-EXIF}), or the ones that are caught by the wake and reside inside it over an extended time interval, capturing the small-scale flow perturbations of the wake stagnation zone and/or its tail which oscillates as vortex shedding occurs. While one can argue that shorter tracks could be used for temporal and geometric statistics acquisition, as well as local PTV, we would like to show how the amount of very long, not just any tracks, changes with $Re_c$, since it is the longer tracks that encode much more information about the spatial and temporal character of turbulent flow and better capture the lower oscillation frequencies.

\begin{figure}[h]
\centering
\includegraphics[width=0.60\linewidth]{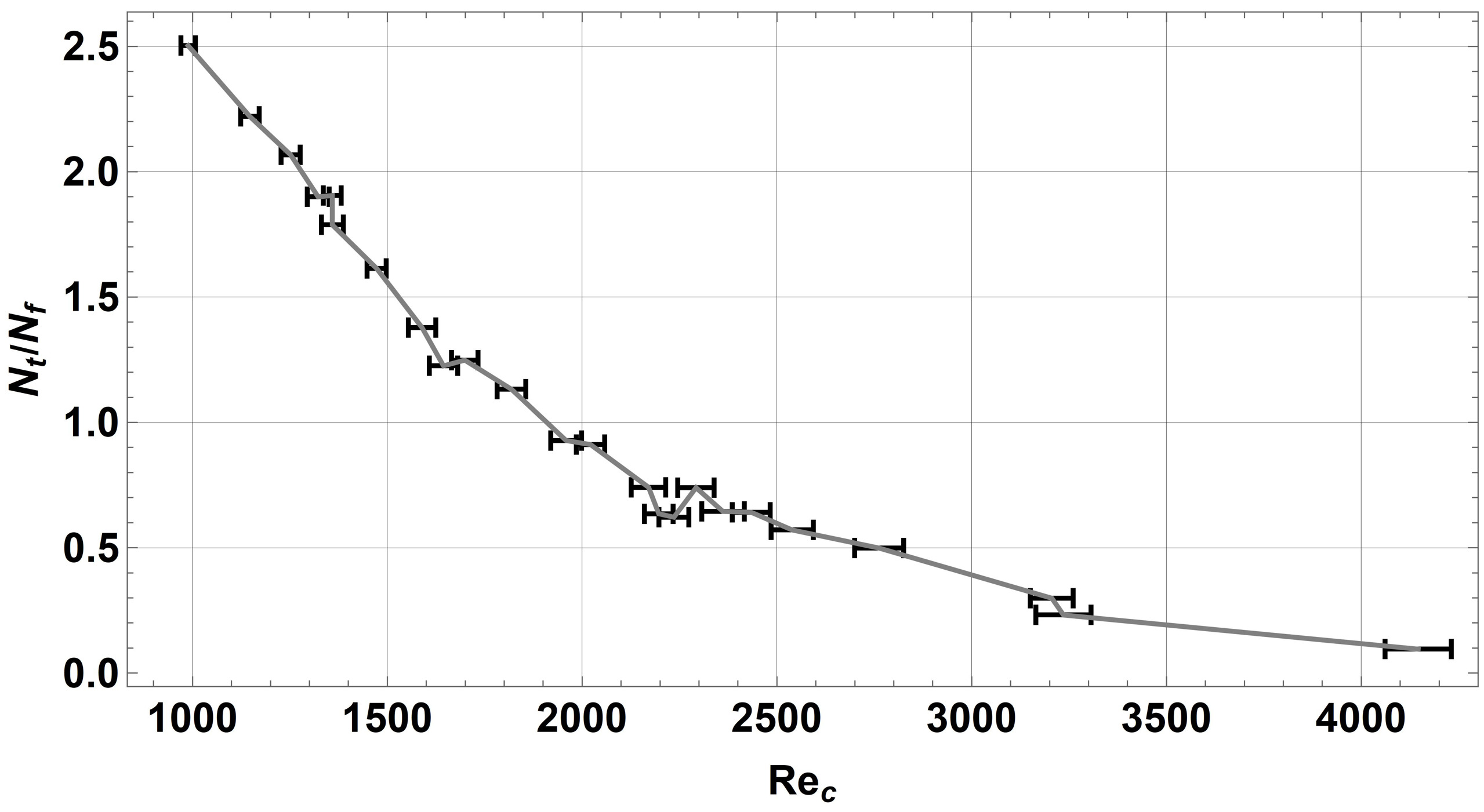}
\caption{Tracking performance of MHT-X: the ratio of total eligible trajectories $N_t$ to the number of frames $N_f$ for a range of $Re_c$.}
\label{fig:tracking-performance}
\end{figure}

As a measure of performance, we use the ratio of the number of eligible trajectories $N_t$ to the number of frames $N_f$ in an image sequence. Figure \ref{fig:tracking-performance} shows how this ratio changes with increasing $Re_c$. Note that $N_t$ in our case typically represents $5\%$ to $10\%$ of the data (on average several thousand trajectories), but this is because the MHT-X outputs trajectories of all lengths, including 1-node trajectories which in the MHT-X framework should be interpreted as eliminated false positives. The eligible track percentages are much greater if tracks with $<5$ nodes are discarded. Observe that the performance degrades at a non-constant rate: we find that $N_t$/$N_f$ fits $x^n$ with $n=-2.13 \pm 0.11$ and $R^2 = 0.94$, but does not fit an exponential function ($R^2 \sim 10^{-3}$) which one might expect if the performance was mostly tied to the velocity-dependent exponential scaling factors in our motion model equations \cite{birjukovs-particle-EXIF}. To improve tracking for higher $Re_c$, one could try and circumvent this by making the respective motion models less restrictive, i.e. allow greater angle changes for faster motion. However, clearly cannot be done indefinitely, since beyond certain $Re_c$ values one will simply require a higher frame rate.

\clearpage

\begin{figure}[h]
\centering
\includegraphics[width=1.00\linewidth]{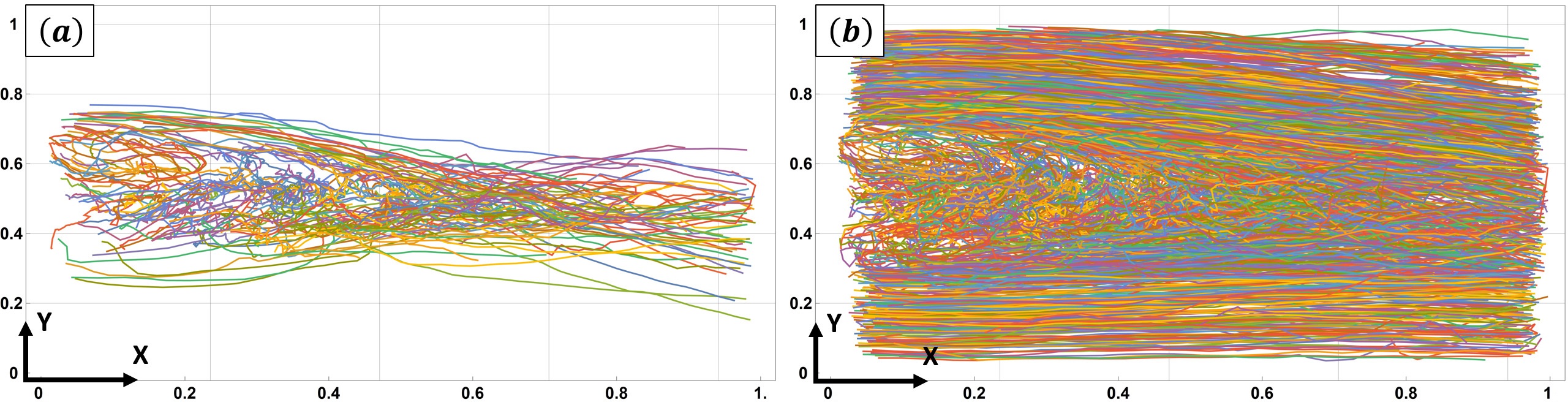}
\caption{Examples of eligible trajectories reconstructed within the FOV: (a) the longest 150 and (b) 1500 trajectories. Colors represent different tracks and axes tick values are normalized to the respective image dimensions.}
\label{fig:eligible-trajectories-example}
\end{figure}

Figure \ref{fig:eligible-trajectories-example} shows representative eligible trajectories. Note that they capture particle motion in the wake stagnation zone, in the wake tail, fly-by particle interactions with the wake, as well as the particles that are mostly unaffected. The amount and node count, as well as the physical lengths of eligible tracks in general by far exceeds the best we could previously achieve \cite{birjukovs-particle-EXIF}. While improved artifact detection and cleanup play an important role, most of the improvement can be attributed to a more accurate PIV-assisted motion prediction stemming from a much better interpolated PIV field yielded by DFI.

\subsection{Particle tracking velocimetry}

\begin{figure}[h]
\centering
\includegraphics[width=0.75\linewidth]{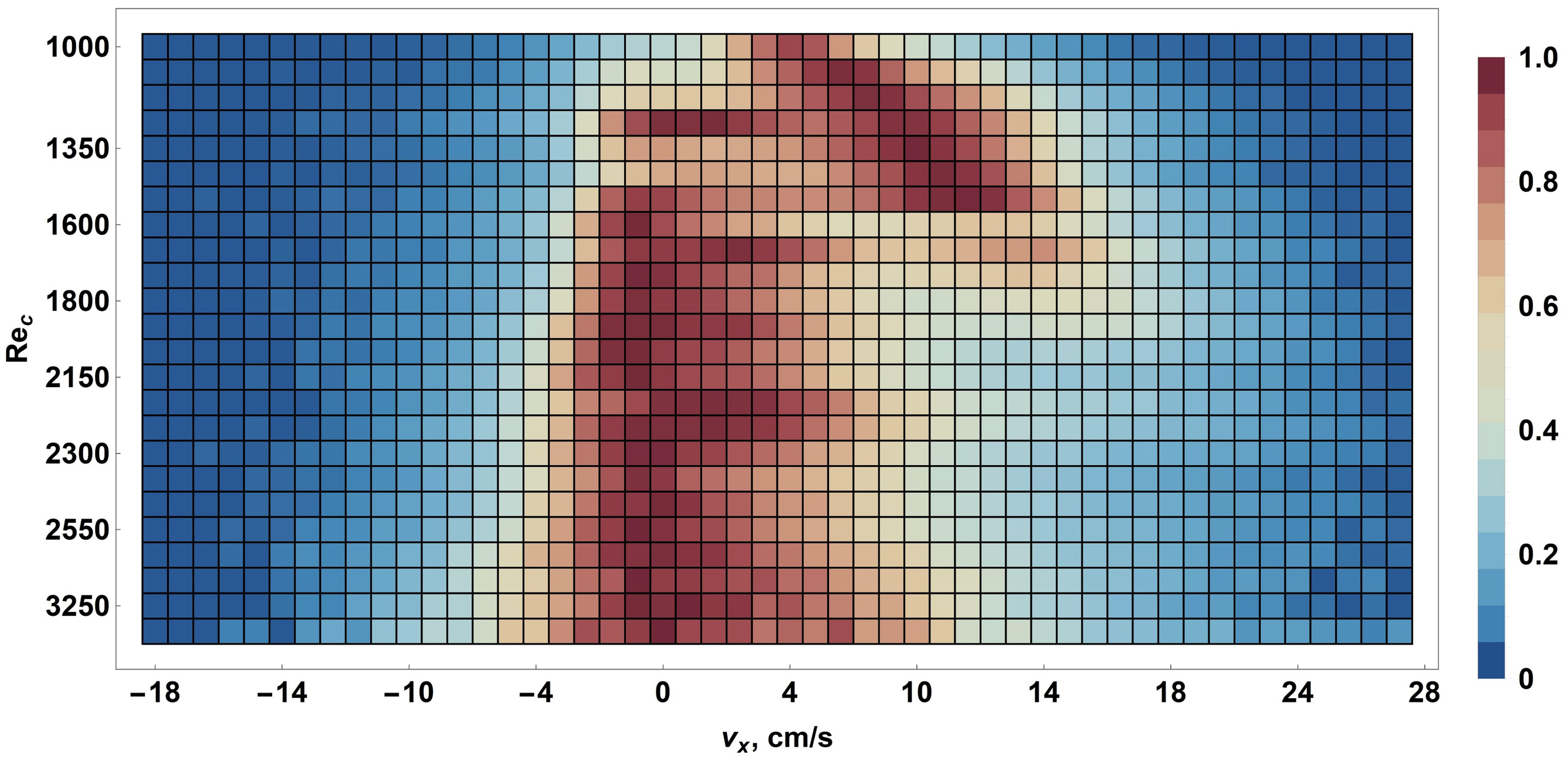}
\caption{Separately binned (Freedman-Diaconis method) and normalized PDFs for the streamwise velocity component $v_x$ of eligible trajectories for a range of $Re_c$.}
\label{fig:velocity-pdf-streamwise}
\end{figure}

\begin{figure}[h]
\centering
\includegraphics[width=1\linewidth]{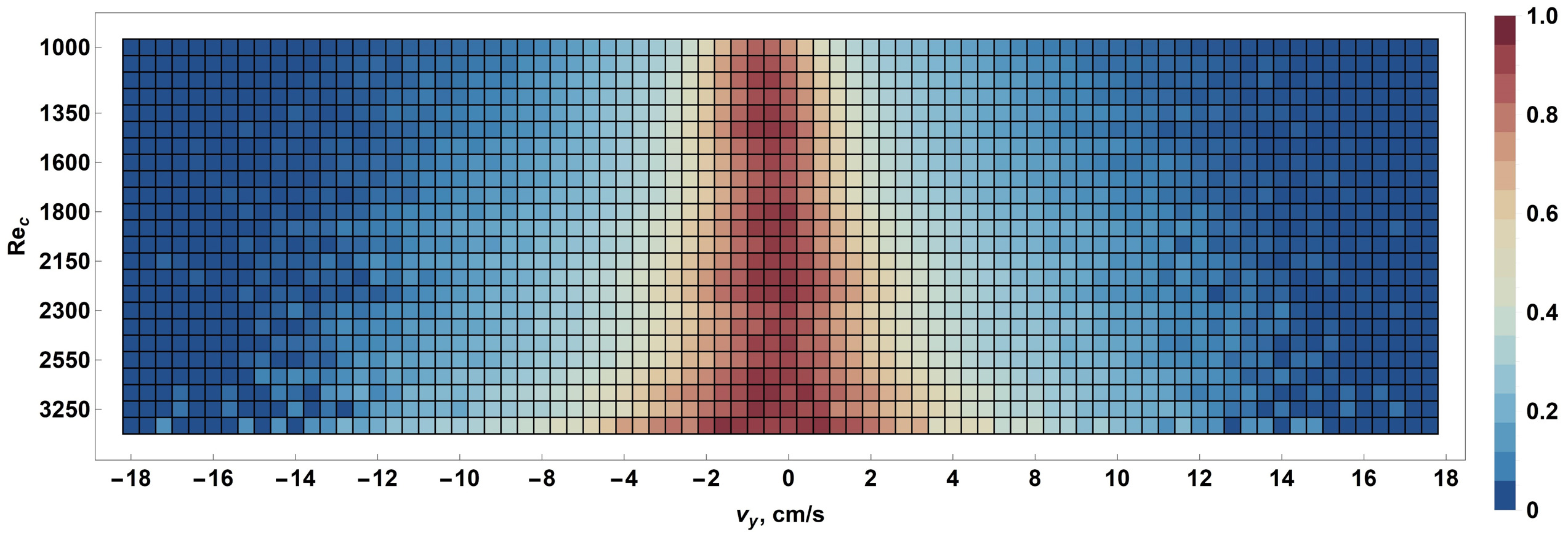}
\caption{Separately binned and normalized PDFs for the transverse velocity component $v_y$ of eligible trajectories for a range of $Re_c$.}
\label{fig:velocity-pdf-transverse}
\end{figure}

As before, we can measure particle displacements between frames, but this time we have enough good trajectories to accumulate particle velocity statistics -- Figures \ref{fig:velocity-pdf-streamwise} and \ref{fig:velocity-pdf-transverse} show the PDF for streamwise $v_x$ and transverse $v_y$ (axes as in Figure \ref{fig:eligible-trajectories-example}) velocity components for all eligible trajectories for each $Re_c$. Of course, as the tracking performance degrades with $Re_c$ approaching its maximum, the PDFs become less smooth and more noisy, and in the case with $Re_c = 4147$ some PDF bins have much smaller weights than they should have (not shown in Figures \ref{fig:velocity-pdf-streamwise} and \ref{fig:velocity-pdf-transverse}). To assess the physical changes in the PDFs as $Re_c$ increases, one could think of the PDFs this way. Particles can be, very roughly, divided into two classes: the ones that are trapped or simply spend a lot of time within the obstacle wake (\textit{captured particles}), and the ones that either pass by unaffected or only weakly interact with the wake (\textit{free particles}), the latter group being faster than the former. One can then expect two things -- first, as $Re_c$ increases, more of the would be free particles will become captured and will have much lower velocities with a distribution centered at near zero velocity, with the velocity dispersion increasing as $Re_c$ does; second, the free particles that remain such will have greater velocity overall and also greater velocity dispersion within their class.

\clearpage

This means that one should expect, at very low $Re_c$, that there only be one maximum in the $v_x$ PDF, associated with the free particles, since the wake will not have developed sufficiently yet. As $Re_c$ increases, it is expected that the initially solitary PDF maximum will begin to split in two as the wake begins to trap particles (the stagnation zone and the tail expand) and the class of captured particles becomes distinguishable. At higher $Re_c$ the residence time of particles captured by the wake increases greatly and the corresponding PDF maximum should begin to dominate the free particle class. At the same time, velocity dispersion about both maxima should increase and the free particle maximum should shift towards greater velocity values. However, since tracking performance degrades with $Re_c$ (Figure \ref{fig:tracking-performance}), one would also expect that the free particle maximum and its neighborhood within the PDF become fainter, since more and more of the faster particles are not properly connected into longer tracks that pass the eligibility criterion. Figure \ref{fig:velocity-pdf-streamwise} exhibits all of the above trends. The splitting of particles into two classes cannot be seen in greater detail because the $Re_c$ sampling is not very fine in the $Re_c \in~ \sim [1000;1200]$ interval, but the transition from free to captured particle domination in the PDFs can be seen fairly clearly in the $Re_c \in~ \sim [1300;1800]$ range. The dispersion of the captured particle velocity is asymmetric with a positive $v_y$ bias that becomes more prominent with increasing $Re_c$ since particles enter and leave the wake zone at higher velocities overall.

As expected, Figure \ref{fig:velocity-pdf-transverse} shows an increased dispersion for the $v_y$ PDF about zero velocity, which is both due to the more intense velocity field fluctuations in the wake stagnation zones, and the stronger oscillations with higher frequency in the wake tail. The fact that this and the above mentioned expected trends are observed suggests that particle detection and tracking are performed adequately.

\subsection{Determining vortex shedding frequencies}

\begin{figure}[h]
\centering
\includegraphics[width=1.00\linewidth]{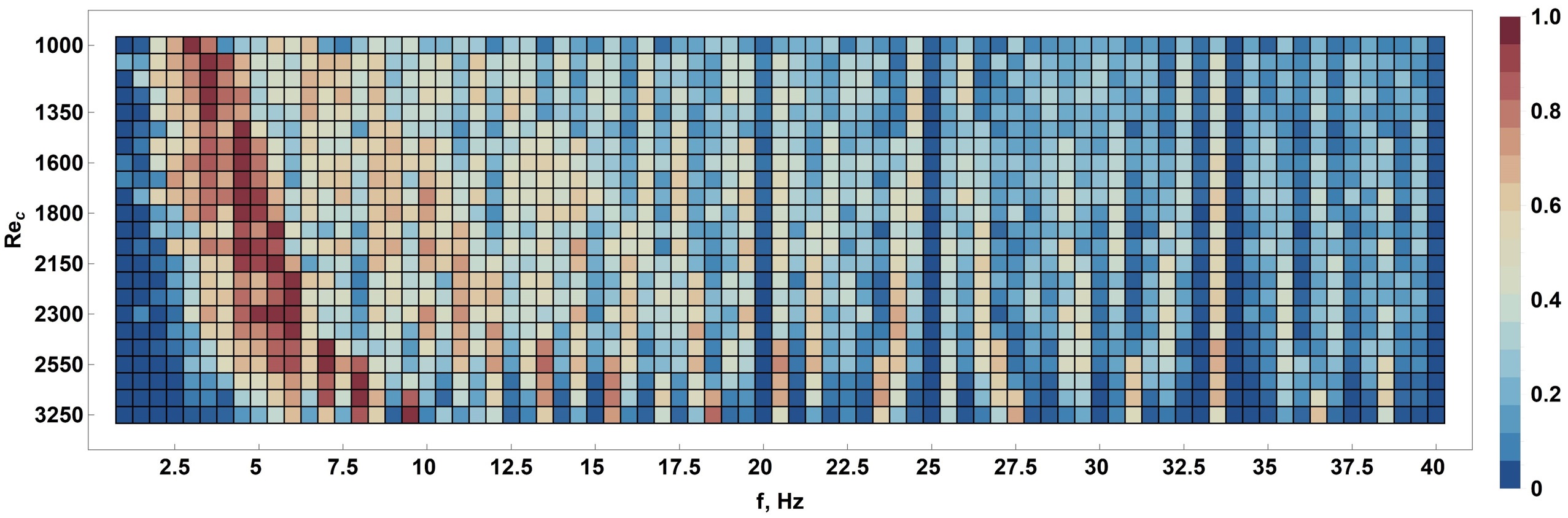}
\caption{Frequency $f$ PDFs for $v_x$ over a range of $Re_c$, aggregated from $v_x$ Fourier spectra of all eligible trajectories in each case. PDF normalization is separate for every $Re_c$ value. Aggregation is done by binning all $f$ instances over the eligible trajectories using Forier coefficients as weights with a 200-bin resolution.}
\label{fig:aggregated-fft-velocity}
\end{figure}

However, this statement still requires further proof and reinforcement. Given the PTV data generated from the eligible trajectories, one can also check if the vortex shedding frequency $f_0$ of the obstacle wake is captured by MHT-X properly -- this is important if one plans to use our methods to assess the temporal characteristics of turbulent flow in experiments with liquid metals. Therefore, $f$ spectra are computed for $v_x$ time series (filtered with the median filter with a kernel width of 1 point for outlier removal, i.e. low-pass filtering) of all eligible trajectories for the considered $Re_c$ range. Then, for each $Re_c$, the $f$ spectra of all trajectories are aggregated by binning all detected $f$ instances using the absolute values of the Fourier coefficients as weights. Doing this for the entire $Re_c$ range and individually normalizing the resulting $f$ PDFs, one obtains Figure \ref{fig:aggregated-fft-velocity}.

\begin{figure}[h]
\centering
\includegraphics[width=0.725\linewidth]{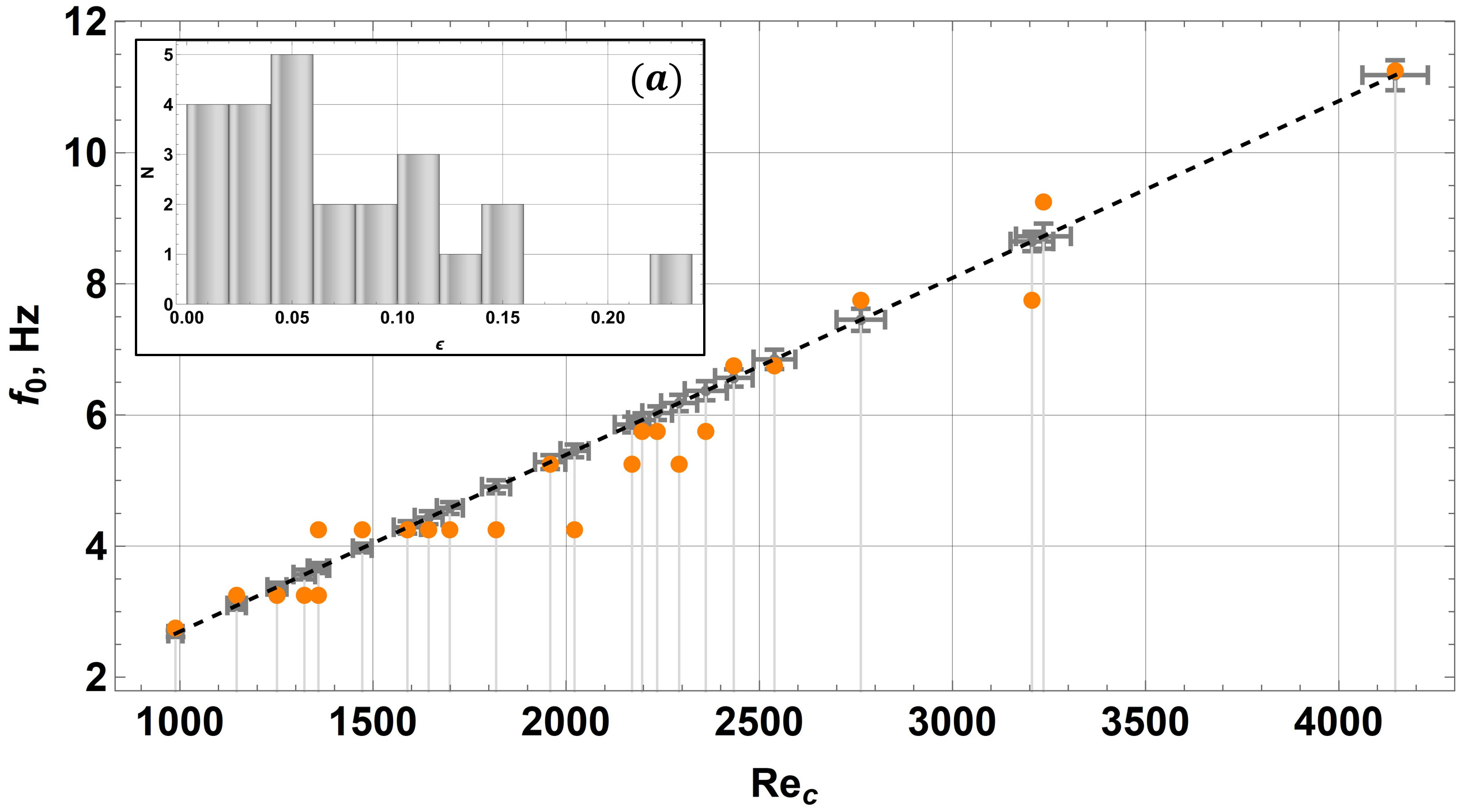}
\caption{Vortex shedding frequency $f_0$: theoretically expected (gray dots with error bars) and derived from the obtained aggregated frequency PDFs (orange dots) by picking the dominant frequency peak for each $Re_c$. The black dashed line indicates the expected linear $f(Re_c)$ trend and the inset (a) shows the histogram for the relative errors ($\epsilon$) between theoretical and experimental $f_0$ values.}
\label{fig:vortex-shedding-expectations-vs-reality}
\end{figure}

Notice the lower-$f$ band with a strong maximum that shifts towards greater $f$ values as $Re_c$ is increased -- this corresponds to the vortex shedding frequency $f_0$. Since we sampled the $Re_c$ range non-uniformly during the imaging experiments, and there are some $Re_c$ instance with rather similar values, one cannot see a clear diagonal in Figure \ref{fig:aggregated-fft-velocity}. However, since the observed shifting band produces the most intense maxima, one can take the maximum PDF values for each $Re_c$, get the corresponding $f$, and plot the values versus $Re_c$, comparing the experimentally determined $f_0$ against what one would expect theoretically. This comparison is shown in Figure \ref{fig:vortex-shedding-expectations-vs-reality} where the expected $f$ values were computed from the $Re_c$ range assuming a constant Strouhal number $Sr = 0.197$ (for the relevant $Re_c$ range, the difference between the minimum and maximum $Sr = 0.198 \cdot (1 - 1.97/Re_\text{c})$ is $\sim 1.5 \%$, so we just take the maximum value). Both visually and upon inspecting inset (a) which shows the relative error histogram for the $Re_c$ range, one can see that the agreement is indeed sound. However, we ask that the reader take the good $f$ match for the maximum $Re_c$ value as an accident, since the corresponding $f$ PDF (not shown in Figure \ref{fig:aggregated-fft-velocity}) has an extremely low signal-to-noise ratio (SNR). The $f_0$ peak for the penultimate $Re_c$ value, however, is very clear despite the objectively worse tracking performance as opposed to the lower $Re_c$ cases. One can also check the quality of $f_0$ determination with a linear fit of the experimentally derived $f_0$. For a constant $Sr$, one has $f_0 = St U_0 / d$ where $U_0$ is the free-stream velocity and $d$ is the obstacle diameter; one also has $Re_\text{c} = \rho U_0 d_\text{c} / \mu$ where $\rho = 6160~ kg/m^3$ is liquid metal density and $\mu = 2.1~ mPa \cdot s$ is its viscosity, meaning $f_0 = Re_c \cdot St / \rho d^2$. Fitting experimental $f_0$ values one gets $Sr = 0.196 \pm 0.009$ which agrees with the expectations.

\subsection{Trajectory curvature statistics}

\begin{figure}[H]
\centering
\includegraphics[width=0.95\linewidth]{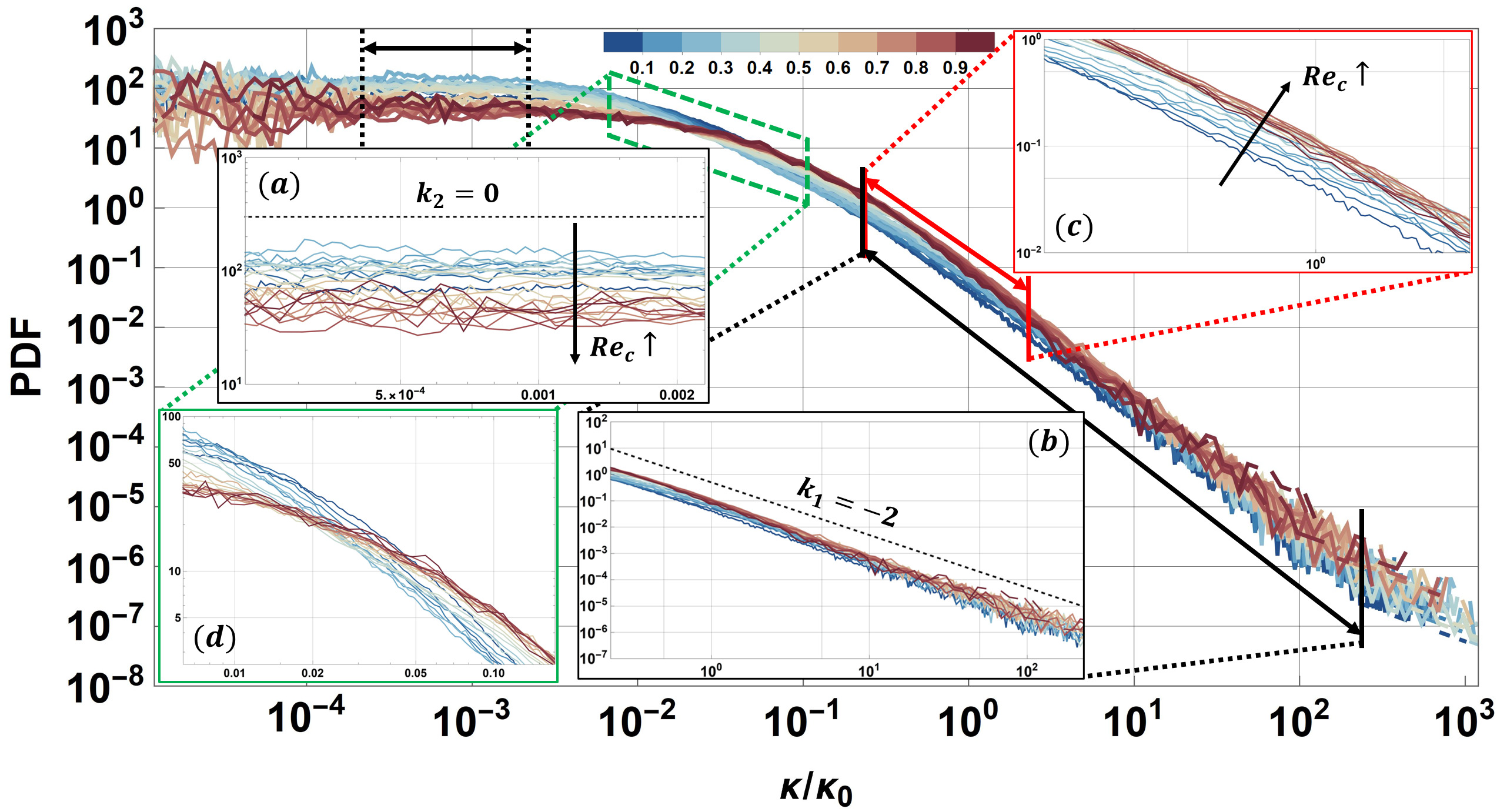}
\caption{Trajectory curvature $\kappa$ PDFs (normalized to the inverse mean particle size $\kappa_0$) for all eligible trajectories for different $Re_c$ with the curves color-coded by their respective $Re_c$ values (color bar 0 corresponds to the minimum $Re_c$ and 1 is the maximum), increasing from blue to red. Note the insets (a) and (b) which highlight the constant ($k_2=0$ reference dashed line) and algebraic decay ($k_1=-2$ reference dashed line) intervals of the PDFs, respectively. Inset (a) shows the downwards shift of the PDF curves in the $k_2=0$ interval as the $Re_c$ value increases, and an inverse trend can be seen in inset (c) for the $k_1=-2$ region. Notice also the trend reversal region highlighted with a green dashed frame and shown in greater detail in inset (d).}
\label{fig:curvature-pdfs}
\end{figure}

In addition to temporal information, spatial characterization of turbulence is also of great interest for liquid metals, therefore we must find out if the quality of trajectory reconstruction by MHT-X enables one to derive trajectory geometry statistics with sufficient accuracy. To this end, we compute curvature $\kappa$ along every eligible trajectory, which is done by fitting a second-order B-spline to trajectory points, and then uniformly re-sampling the trajectory from the spline with an up-sampling factor of 15. This leaves even the finer features of trajectories intact while removing sharp corners which would have generated unphysical curvature values. Afterwards, $\kappa$ is measured at every newly generated point. All of the determined curvature values are normalized the the inverse mean particle size ($\kappa_0 = 1/\langle d_\text{p} \rangle$) and a PDF for $\kappa$ is computed in the double-$\log_{10}$ domain using Freedman-Diaconis binning. This is done for the entire $Re_c$ range, yielding the PDFs shown in Figure \ref{fig:curvature-pdfs}.

There are several key features of the PDFs that must be noted. First of all, one can see very clearly, especially in insets (a) and (b), that the PDFs indeed exhibit a near-constant $k_2=-0.05 \pm 0.05$ low-$\kappa$ region (with a much smaller SNR below $\kappa/\kappa_0 \sim 2.5 \cdot 10^{-4}$) and an algebraic decay region with $k_1 = - 2.09 \pm 0.04$, both of which are rather close the the expected values of $k_2 = -0$ and $k_1 = -2$ (although $k_1=-2.1$ and $k_1=-2.07$ were also reported in some relevant cases -- please see Section \ref{sec:intro}). In addition, two trends can be observed: in the low-$\kappa$ interval, the PDF curves shift downwards as $Re_c$ increases, as shown in inset (a) -- this matches what was measured experimentally in \cite{curvature-flow-topology}; for the high-$\kappa$ region, an inverse trend is seen in inset (c) -- this, while not stated, can still be seen in \cite{curvature-flow-topology} despite PDFs shown therein being compressed (without modifying $k_{1,2}$) by scaling the curvature values with respect to velocity and acceleration variations. Finally, one can clearly see the $Re_c$ trend inversion in inset (d), also observed in \cite{curvature-flow-topology}. The fact that we successfully reproduce these elements of the $\kappa$ PDFs and their universal scaling over the $Re_c$ range considered herein serves to further validate the applicability of our methods to liquid metals.

\begin{figure}[h]
\centering
\includegraphics[width=1\linewidth]{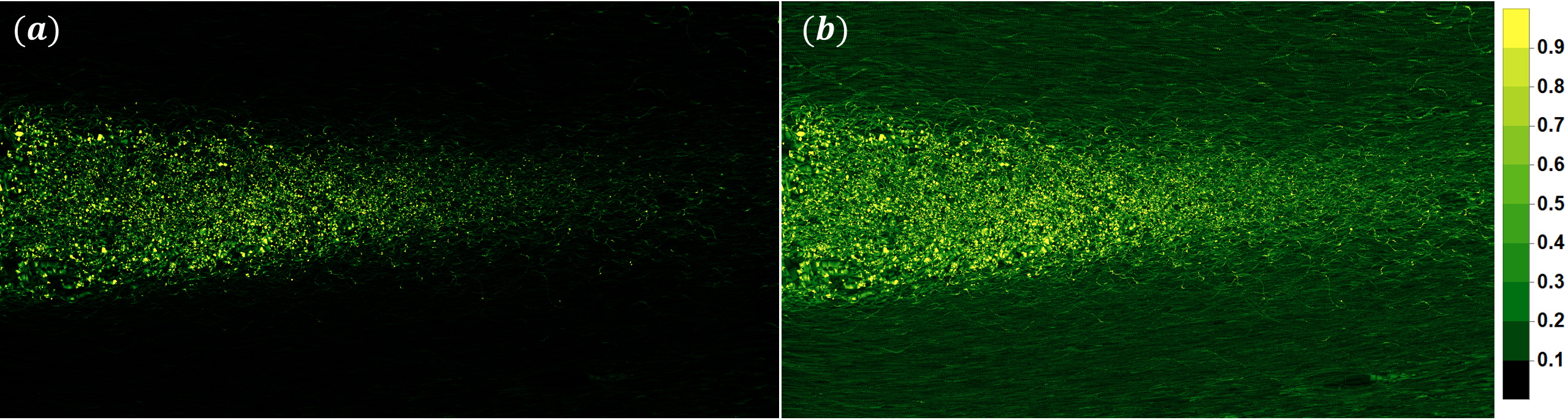}
\caption{An example of a trajectory curvature $\kappa$ map in the imaging FOV obtained from eligible trajectories. (a) Shows the colormap of normalized $\log_{10}(\kappa/\kappa_0)$ and (b) is its version with color tone mapping applied.}
\label{fig:curvature-fov-map}
\end{figure}

However, we also wanted to make sure that it is not just the $\kappa$ PDFs that are computed correctly, but also the $\kappa$ maps over the FOV. To do this, we linearly interpolate the sparse $\kappa$ field aggregated from all eligible trajectories for a given $Re_c$ and then re-sample it with sub-pixel resolution of $0.2~px$ (a trade-off between detail preservation and noise/outlier elimination) -- an example for one of the analyzed image sequences can be seen in Figure \ref{fig:curvature-fov-map} where a color map of $\log_{10}(\kappa/\kappa_0)$ is shown. The higher $\kappa$ values are concentrated within the obstacle wake zone, as expected, with faint narrower filaments extending over to the right side of the FOV. The filament-like structure seen in Figure \ref{fig:curvature-fov-map} can be understood by applying color tone mapping to the $\log_{10}(\kappa/\kappa_0)$ map before colorizing it -- compressing the map value histogram reveals clear curves within and outside of the wake zone that are nothing other than visualized trajectories, where yellower spots and filaments are higher-$\kappa$ fragments of trajectories.

\begin{figure}[H]
\centering
\includegraphics[width=1.00\linewidth]{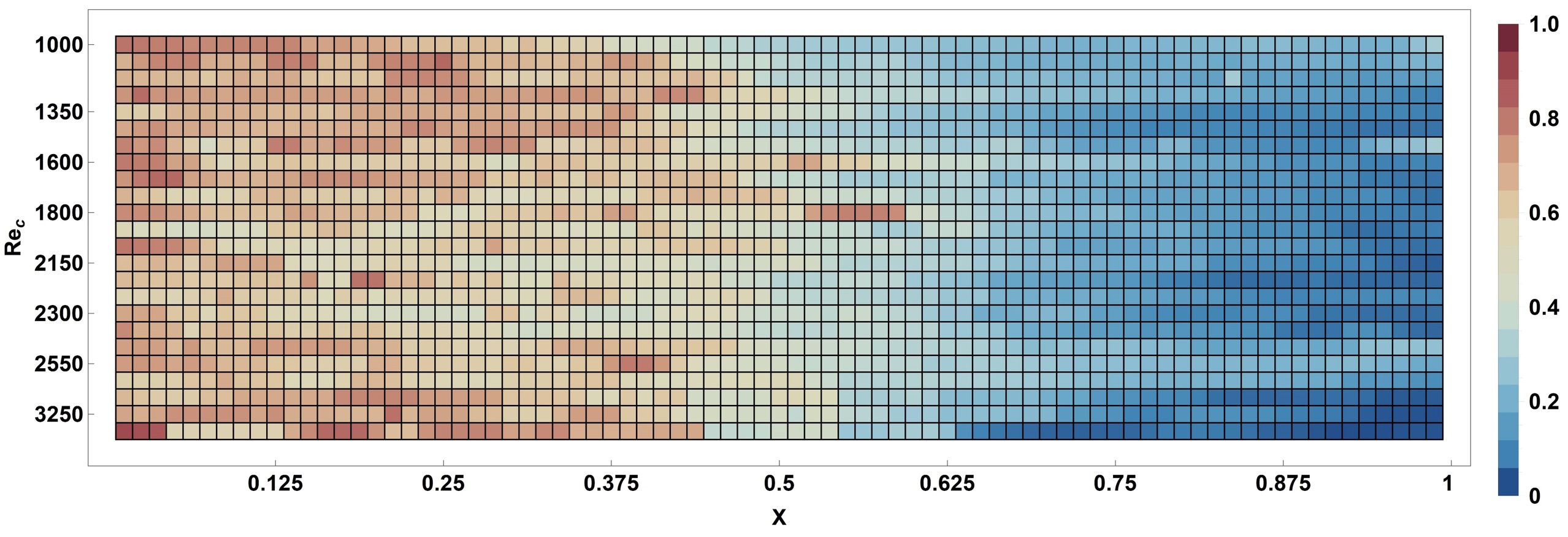}
\caption{Mean curvature profiles over $X$ (streamwise) for the range of $Re_c$. The color map encodes normalized (individually for each $Re_c)$ median filtered (kernel radius is 2 points) $\log_{10}(\kappa/\kappa_0)$.}
\label{fig:curvature-x-profiles}
\end{figure}

\begin{figure}[H]
\centering
\includegraphics[width=0.75\linewidth]{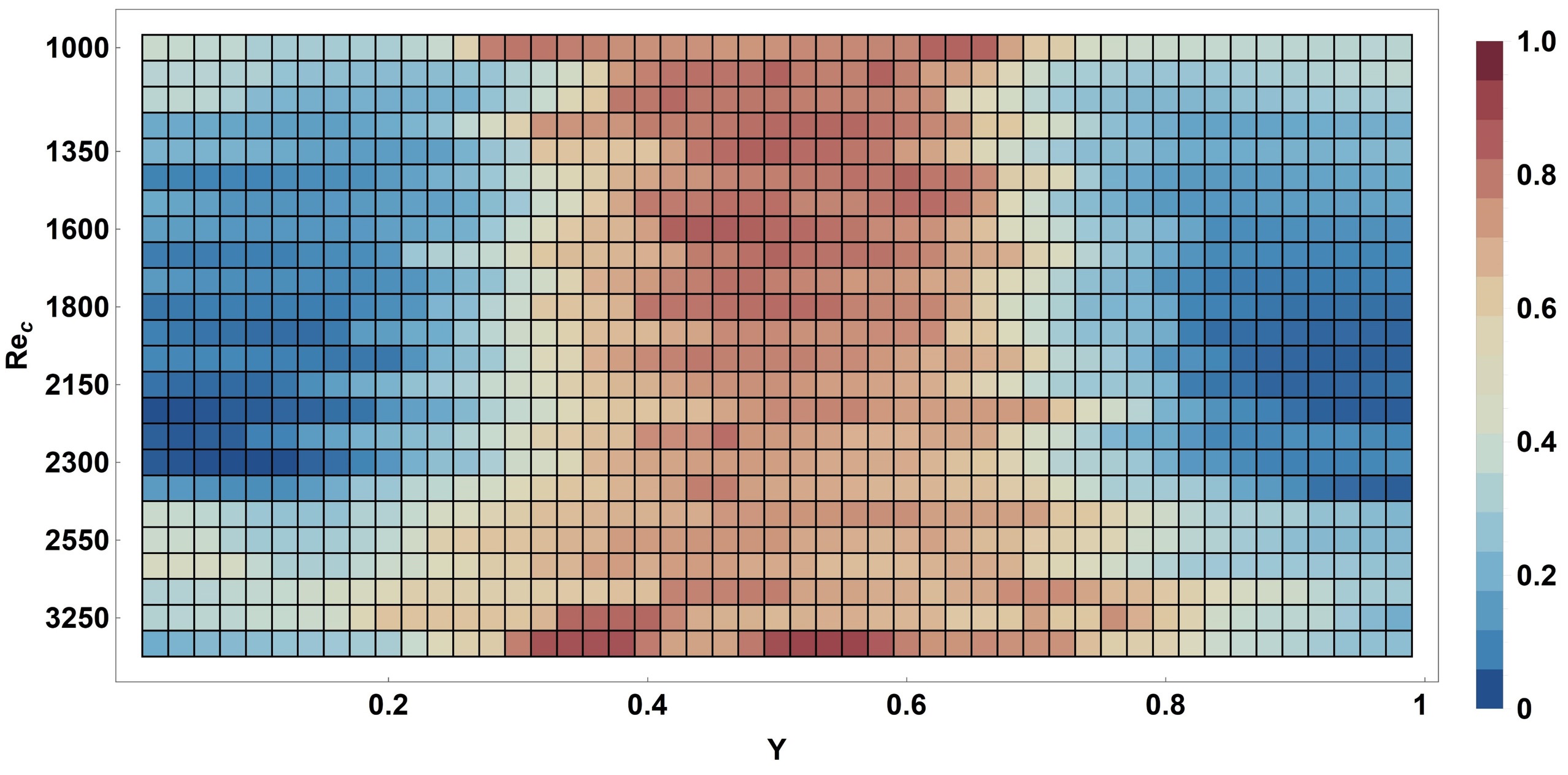}
\caption{Mean curvature profiles over $Y$ (transverse) for the range of $Re_c$. The color map encodes normalized median filtered $\log_{10}(\kappa/\kappa_0)$.}
\label{fig:curvature-y-profiles}
\end{figure}

\begin{figure}[H]
\centering
\includegraphics[width=1.00\linewidth]{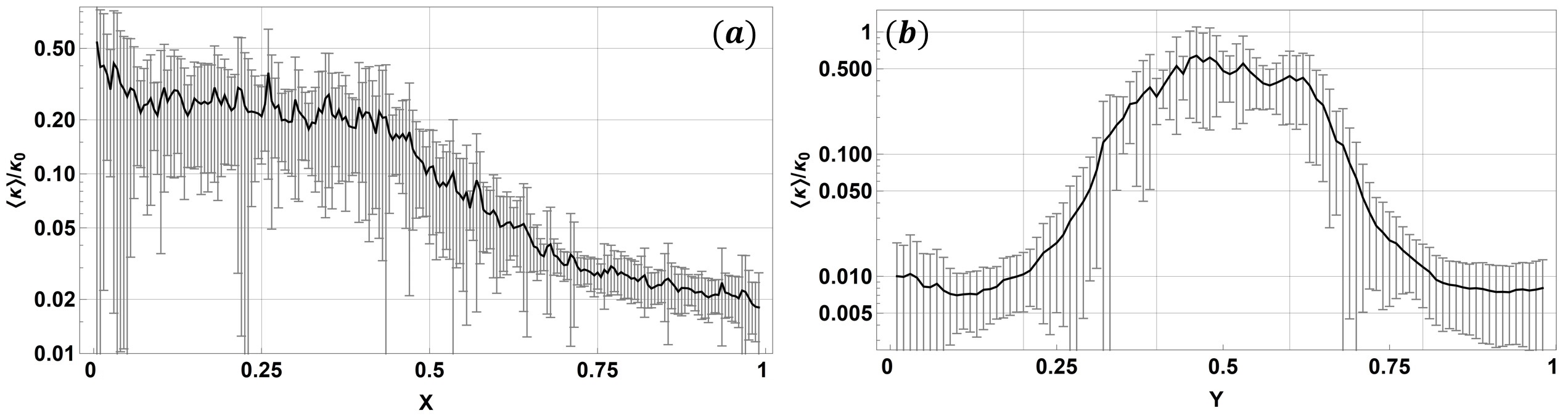}
\caption{Mean $\kappa/\kappa_0$ profiles across the $Re_c$ ranges over (a) $X$ and (b) $Y$ directions in the FOV.}
\label{fig:curvature-mean-profiles}
\end{figure}

Interestingly, the $\kappa$ distribution patterns seen in $\kappa$ maps for different $Re_c$ are rather similar. Figures \ref{fig:curvature-x-profiles} and \ref{fig:curvature-y-profiles}, which are derived from Figure \ref{fig:curvature-fov-map}a by coarser binning, show how mean $\kappa$ profiles over $X$ and $Y$ (axes as in Figure \ref{fig:eligible-trajectories-example}) change with $Re_c$, and Figure \ref{fig:curvature-mean-profiles} shows mean $\kappa$ profiles for all considered $Re_c$ values. Notice that while the $X$ profiles are quite similar, one can see that the $Y$ profiles' width slightly scales with $Re_c$, indicating the the obstacle wake expands transversely as the free-stream velocity increases, which makes sense. Figure \ref{fig:curvature-mean-profiles} reveals that the curvature (in the semi-$\log_{10}$ domain) has a plateau with a very small slope that extends from the edge of the cylindrical obstacle ($X=0$) almost to the middle of the FOV, and then quickly falls of with $X$. The $Y$ profile is, as expected, quite symmetric.

\begin{figure}[h]
\centering
\includegraphics[width=0.60\linewidth]{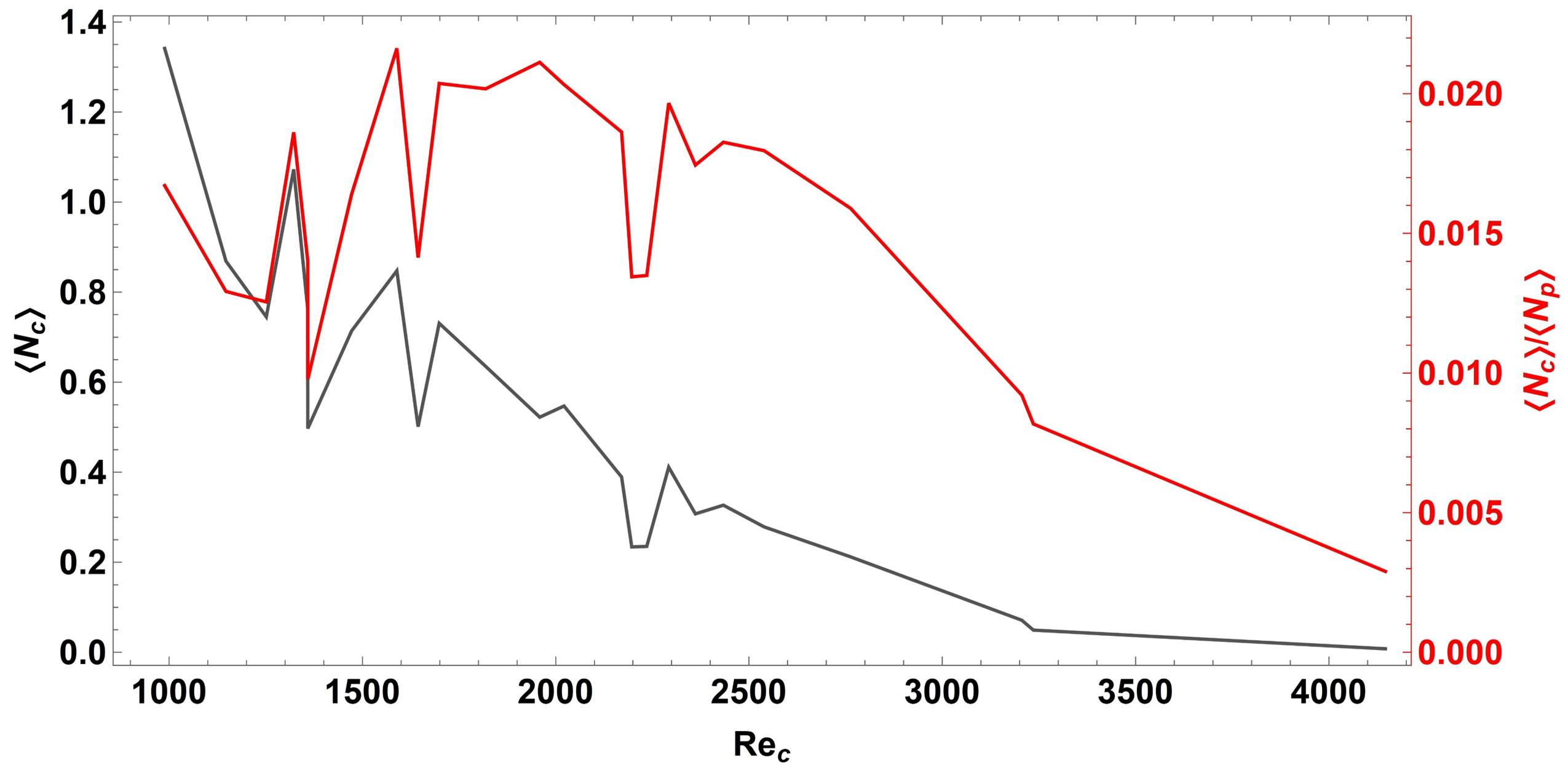}
\caption{Mean particle collision count per frame $\langle N_\text{c} \rangle$ (gray) and its ratio to the mean number of particles per frame $\langle N_\text{p} \rangle$ (red) for the considered $Re_c$ range.}
\label{fig:collision-frequency}
\end{figure}

Of course, another concern one must address is that the $\kappa$ statistics shown in Figure \ref{fig:curvature-pdfs} could be affected by particle collisions. While we do not expect a significant amount of collisions in any given $Re_c$ case, we should provide at least a crude estimate for how much of an effect collisions could have. To do this, we use a primitive collision estimation model which considers particles from eligible trajectories which are simultaneously within a given frame, and then we check their distances and velocity to see if collisions could occur. Specifically, only particles within $r_\text{c} = \langle d_\text{p} \rangle + \sigma_\text{d}$ ($\sigma_\text{d}$ is the particle diameter standard deviation) of one another are considered candidates for a potential collision for a given frame. Whether or not they are actually likely to collide is determined by their relative velocity projected onto a line connecting their centroids -- if the particles are traveling away from one another or if the projected relative velocity is less than $5\%$ of the free-stream velocity, the particles are ruled out as potentially colliding pairs. With this primitive model we can roughly estimate how often the collisions might occur, and where -- this is reflected in Figure \ref{fig:collision-frequency} where the mean particle collision count per frame $\langle N_\text{c} \rangle$ and its ratio to the mean number of particles per frame $\langle N_\text{p} \rangle$ are shown for the considered $Re_c$ range.

\clearpage

\begin{figure}[h]
\centering
\includegraphics[width=0.70\linewidth]{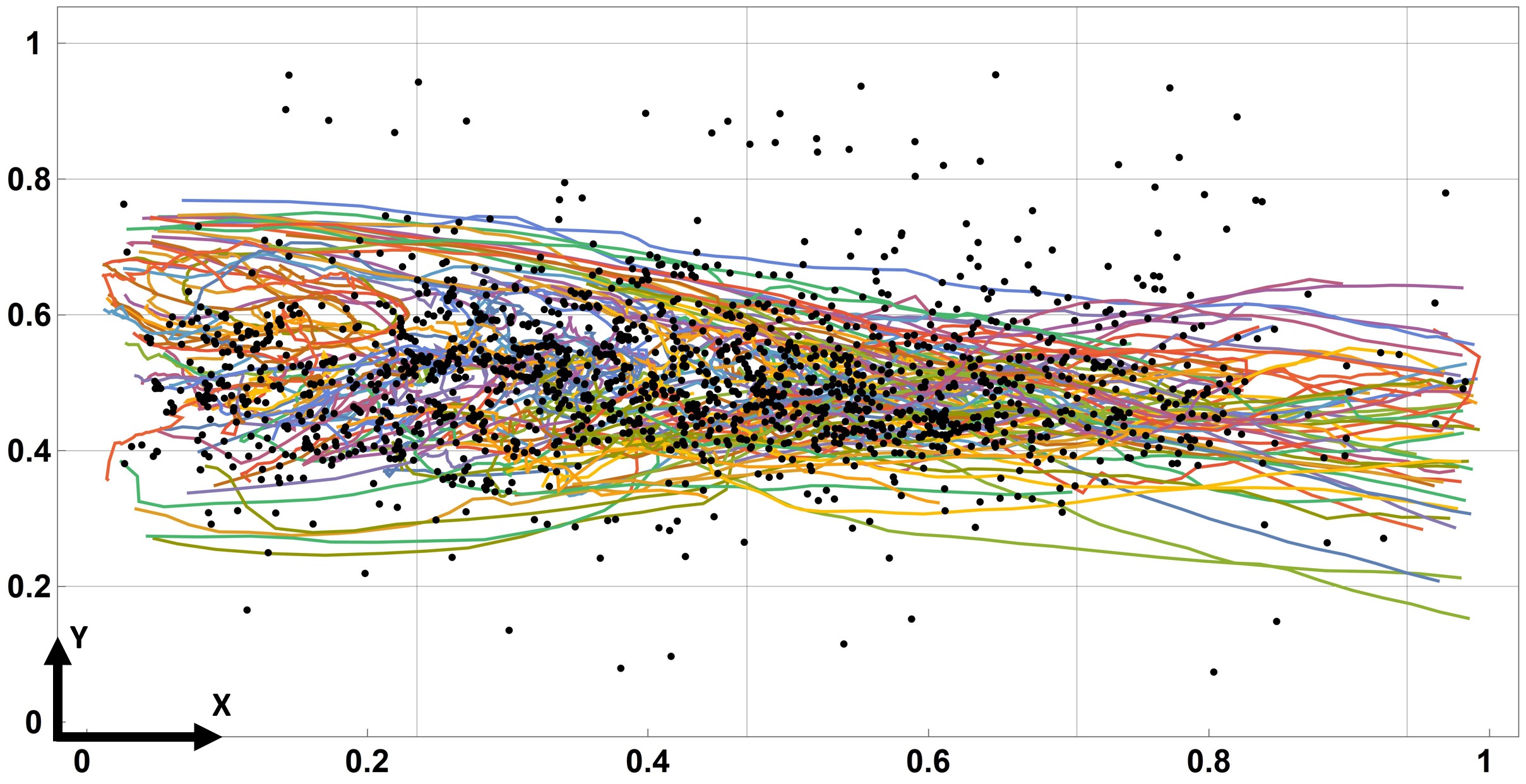}
\caption{An example of collision locations (black dots) estimated within the FOV. Trajectories colored by their IDs are shown for context.}
\label{fig:collision-location-examples}
\end{figure}

We remind the reader that one must also consider the degradation of MHT-X tracking performance with increasing $Re_c$, which explains the rather sharp falloff for $\langle N_\text{c} \rangle/\langle N_\text{p} \rangle$ past the $Re_c \sim 2500$ mark. However, one can see that even for $Re_c < 2500$ where we should be able to see potential collisions with the above simple model, they are quire rare, especially considering that there is on average $\sim 250$ particles within the FOV simultaneously. An example of how collisions estimated for an entire image sequence are distributed within the FOV are shown in Figure \ref{fig:collision-location-examples}.

\begin{figure}[h]
\centering
\includegraphics[width=0.95\linewidth]{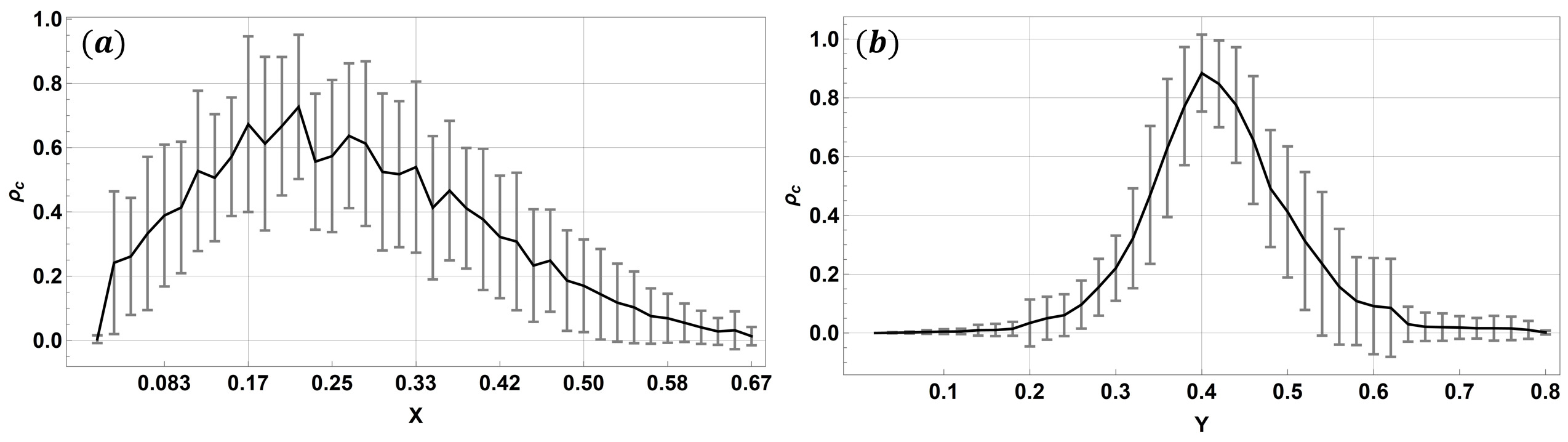}
\caption{Mean collision number density $\rho_c$ over (a) $X$ and (b) $Y$ within the FOV for the considered $Re_c$ range.}
\label{fig:collision-density-mean-profiles}
\end{figure}

It is evident that the overwhelming majority of collisions occur within the wake, particularly in the stagnation zone where particles are subject to sharp and rather chaotic perturbations in the velocity field. This is clear from Figure \ref{fig:collision-density-mean-profiles} where an interval containing the maximum collision number density lies within the stagnation zone seen in Figure \ref{fig:collision-location-examples}. The collision number density profiles are quite consistent between the different $Re_c$ values.

\begin{figure}[H]
\centering
\includegraphics[width=0.925\linewidth]{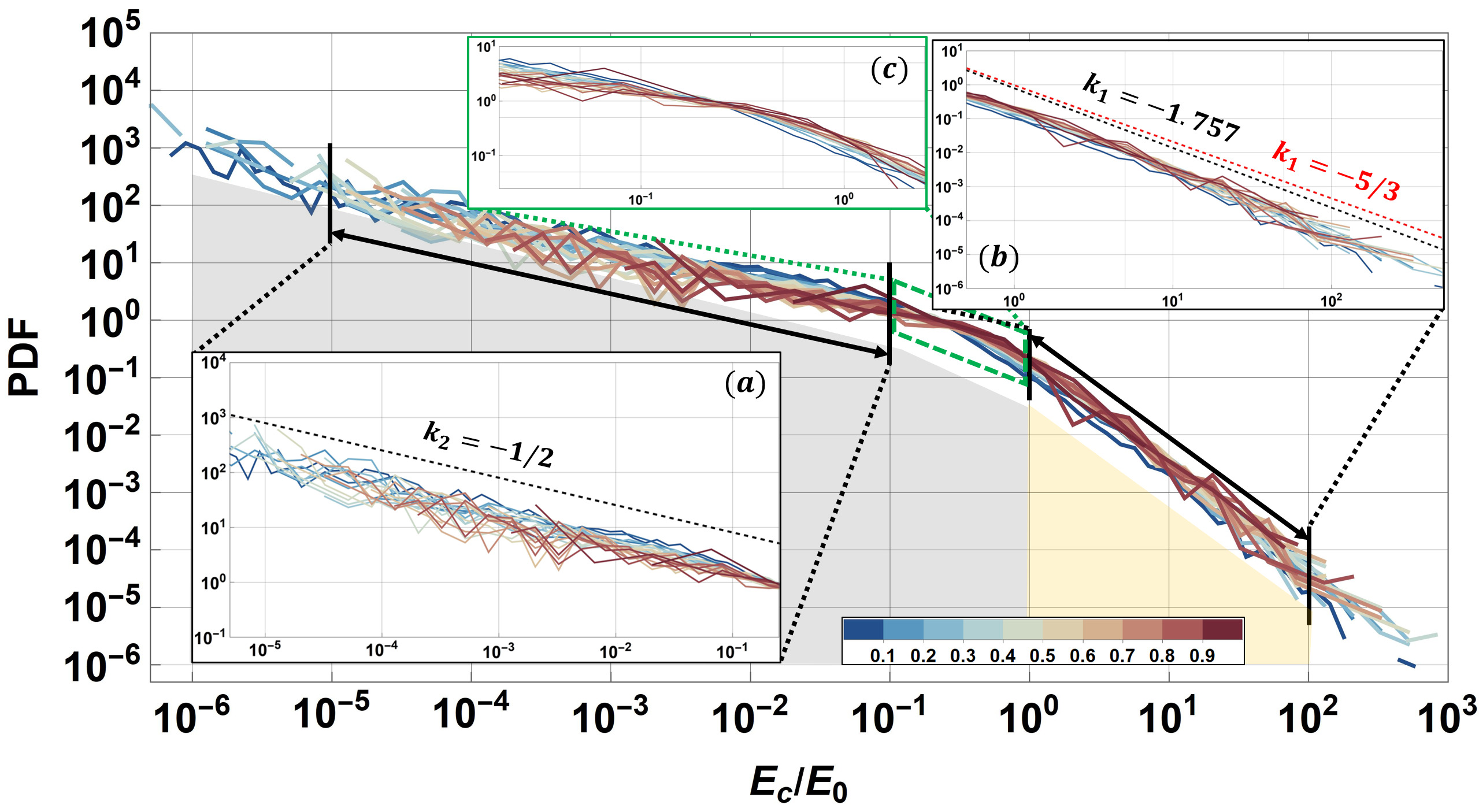}
\caption{Collision to initial kinetic energy ratio $E_\text{c}/E_0$ PDFs for all estimated collisions for different $Re_c$. The curves are color-coded by their respective $Re_c$ values (0 corresponds to the minimum $Re_c$ and 1 is the maximum), increasing from blue to red. Insets (a) and (b) highlight the $k_2 = -1/2$ (reference dashed line) and $k_1=-1.757$/$k_1 = -5/3$ (reference black and red dashed lines) intervals of the $E_\text{c}/E_0$ PDFs, respectively. Inset (c) shows the $Re_c$ PDF shift trend inversion like the one seen in Figure \ref{fig:curvature-pdfs}d. Gray- and yellow-tinted areas represent the cumulative contributions of the low- and high-$E_\text{c}/E_0$ parts of the PDFs, with a boundary at $E_\text{c}/E_0 = 1$.}
\label{fig:collision-energy-pdf}
\end{figure}

Knowing how often and where collisions occur the most, we can estimate their impact on particle motion and trajectories, particularly their $\kappa$ statistics. We can estimate the collision energy $E_\text{c}$ from the relative velocity of colliding particles projected onto the line connecting their centroids. To make it more straightforward yet fair, we assume the worst-case scenario and assign the total $E_\text{c}$ (particle mass times projected relative velocity) to \textit{both} particles of the colliding pair. Since the particle size deviation $\sigma_\text{d}$ is quite small, we also assume that all particles have \textit{the same} mass. While $E_\text{c}$ itself is not informative, its ratio to the initial kinetic energy $E_0$ of each particle in a pair should determine to what relative extent the kinetic energy changes and therefore how affected the trajectory shape can be. Computing $E_\text{c}/E_0$ for every particle estimated to participate in a collision, then constructing a PDF in the double-$\log_{10}$ domain (Freedman-Diaconis binning) and doing this for each $Re_c$ case, one has the $E_\text{c}/E_0$ PDFs as seen in Figure \ref{fig:collision-energy-pdf}.

\begin{figure}[h]
\centering
\includegraphics[width=0.95\linewidth]{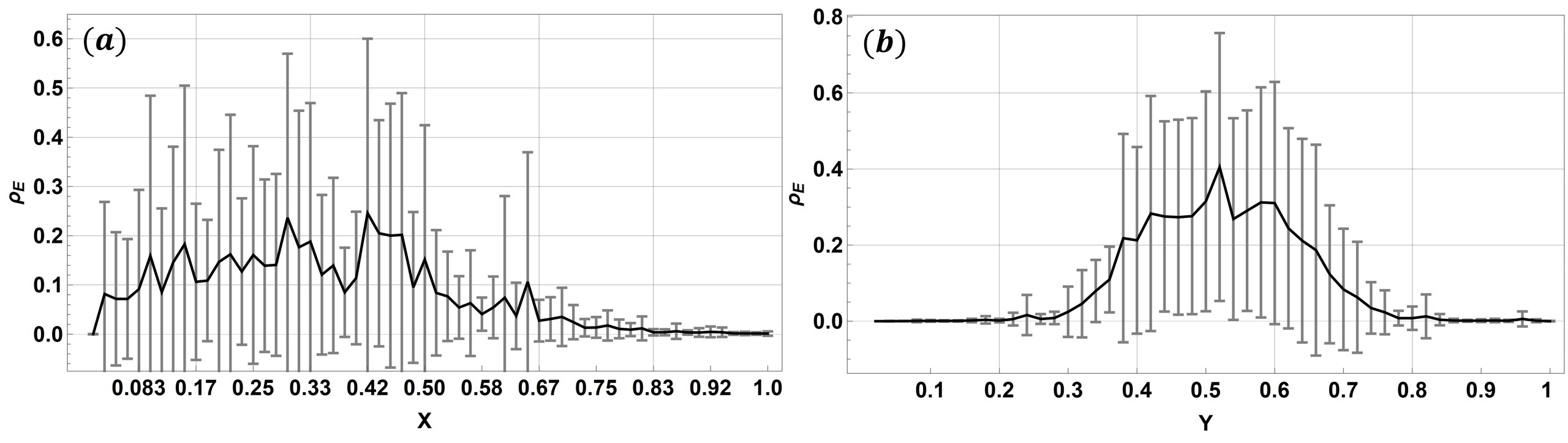}
\caption{Mean normalized $E_\text{c}/E_0$ profiles over (a) $X$ and (b) $Y$ within the FOV for the considered $Re_c$ range.}
\label{fig:collision-energy-mean-profiles}
\end{figure}

\begin{figure}[H]
\centering
\includegraphics[width=0.775\linewidth]{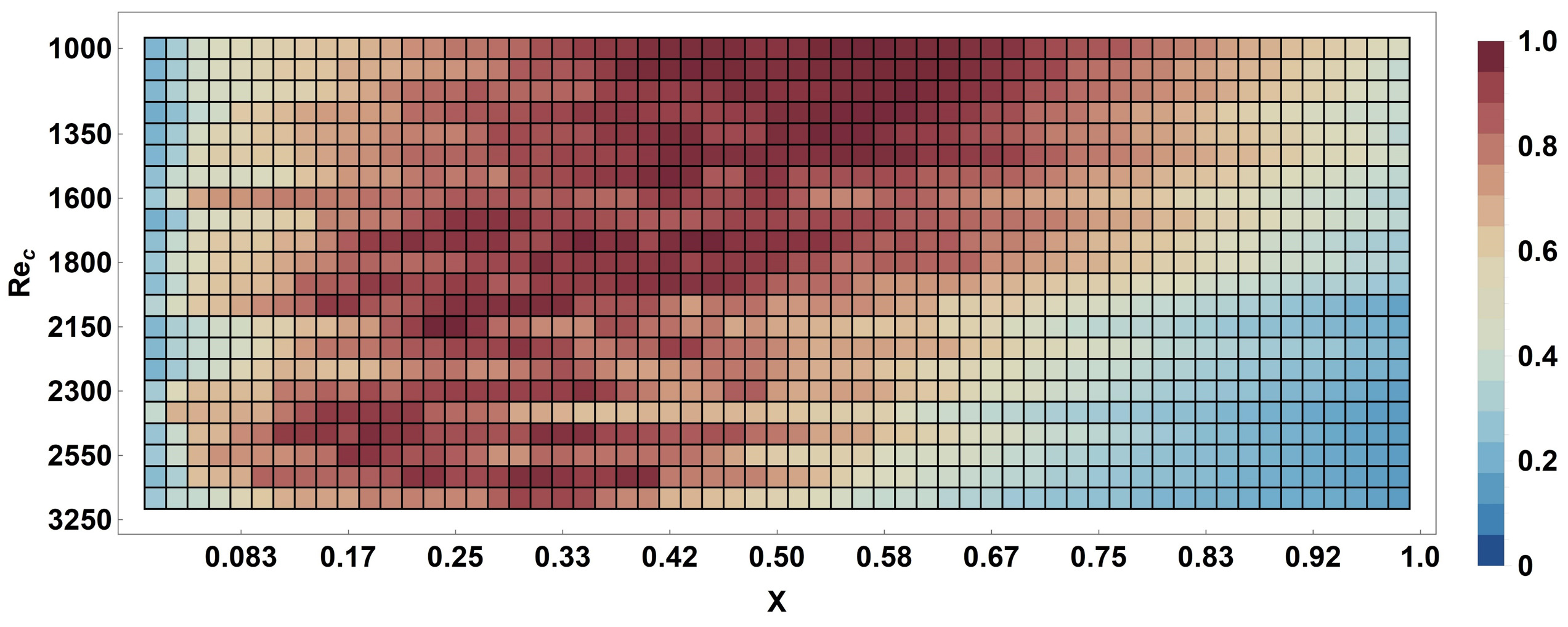}
\caption{Mean particle number density $\rho_\text{c}$ over $X$ for the range of $Re_c$. The color map encodes normalized (individually for each $Re_c)$ median filtered (kernel radius is 1 point) $\rho_\text{c}$.}
\label{fig:particle-density-x-profiles}
\end{figure}

\begin{figure}[H]
\centering
\includegraphics[width=0.65\linewidth]{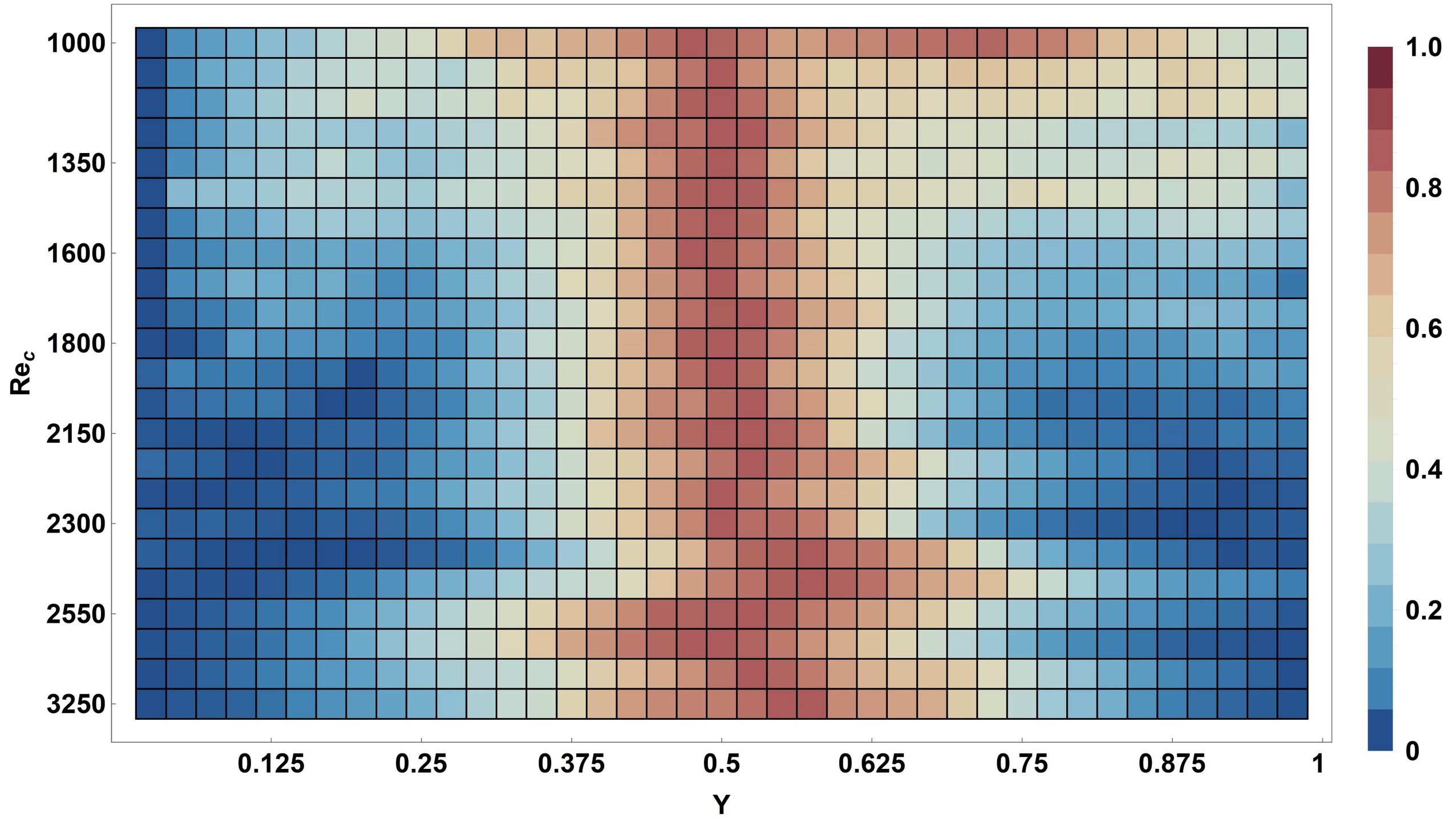}
\caption{Mean particle number density $\rho_\text{c}$ over $Y$ for the range of $Re_c$. The color map encodes normalized median filtered $\rho_\text{c}$.}
\label{fig:particle-density-y-profiles}
\end{figure}

\begin{figure}[h]
\centering
\includegraphics[width=0.95\linewidth]{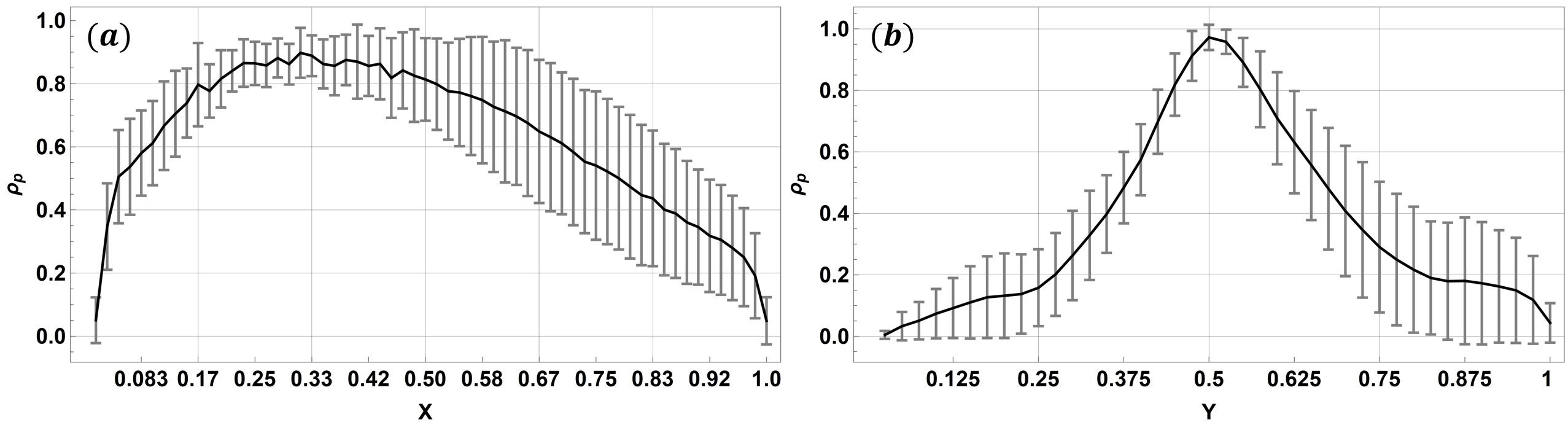}
\caption{Mean normalized particle number density $\rho_\text{p}$ profiles over (a) $X$ and (b) $Y$ within the FOV for the considered $Re_c$ range.}
\label{fig:particle-density-mean-profiles}
\end{figure}

The $E_\text{c}/E_0$ PDFs also exhibit algebraic decay and two distinct regions (high- and low-ratio) -- but unlike the $\kappa/\kappa_0$ PDFs in Figure \ref{fig:curvature-pdfs}, algebraic decay occurs in both of the regions. It is tempting to say that insets (a) and (b) show PDF shifting with $Re_c$ like what is seen for $\kappa/\kappa_0$ in Figure \ref{fig:curvature-pdfs}, but the SNR here is insufficient for us to state this with confidence anywhere except perhaps the $E_\text{c}/E_0 \in ~ \sim (10^{-2};10^{-1})$ interval. However, the $Re_c$ PDF shift trend inversion similar to that in Figure \ref{fig:curvature-pdfs}d is rather clear in inset (c). As for the algebraic decay intervals, here one has a low-ratio exponent $k_2 = -0.498 \pm 0.009$ (almost $1/2$, although the fit goodness is not so good with $R^2 = 0.89$) and a high-ratio exponent $k_1 = -1.757 \pm 0.011$, the latter being very close to $-5/3$, which is a very interesting coincidence (but, as far as we can tell, just that). Notice that the transition between the $k_2$ and $k_1$ algebraic decay exponents occurs very close to $E_\text{c}/E_0 = 1$, starting at $\sim 10^{-1}$ which is, roughly speaking, the threshold beyond which collisions should start noticeably affecting particle trajectories. However, regarding the initial question of just how significant the collisions could be overall, it is abundantly clear from Figure \ref{fig:collision-energy-pdf} that the absolutely dominant contribution to the respective cumulative distribution function stems from low energy ratio collisions. This means that the $\kappa$ statistics derived from the tracks should indeed encode mostly the effects of turbulent pulsations within the liquid metal flow. Because $\langle N_\text{c} \rangle/\langle N_\text{p} \rangle$ is very low, and there is a very wide range of $E_\text{c}/E_0$ values, the SNR of the $E_\text{c}/E_0$ profiles for individual $Re_c$ is also not high enough to determine where the $E_\text{c}/E_0$ maxima are in the FOV within acceptable error margins, as seen in Figure \ref{fig:collision-energy-mean-profiles}. To do this, especially for greater $Re_c$, significantly longer imaging time in required per image sequence. Finally, if one wishes to study collisions in-depth, a rigorous approach is required and ideally collisions should be built into the MHT-X motion models. We can, however, assess where in the FOV the particles spend more time, which is shown in Figures \ref{fig:particle-density-x-profiles}-\ref{fig:particle-density-mean-profiles}.

Figure \ref{fig:particle-density-x-profiles} suggests that, as $Re_c$ increases, more and more particles are trapped within the wake, and specifically its stagnation zone, therefore the maximum of the particle density ($\rho_\text{p}$) profile shifts towards smaller $X$. Averaging over $Re_c$ one also find that the stagnation zone is the most particle-populated area of the FOV. Since the frame rate is constant, $\rho_\text{p}$ corresponds to the particle residence time. However, note that due to tracking performance degradation for greater $Re_c$, some of the particles that travel with faster velocity and only weakly interact with the wake tail will be lost. This is why the $\rho_\text{p}$ profiles for higher $Re_c$ do not extend as much beyond the $X=0.6$ mark. Figure \ref{fig:particle-density-y-profiles}, as expected, implies that with increasing $Re_c$ the relative particle residence time, which $\rho_\text{p}$ represents, becomes more at the center ($Y=0.5$) and less away from it, with greater dispersion for higher $Re_c$ due to wake width increase.

\subsection{Velocity field reconstruction}

\begin{figure}[H]
    \centering
    \includegraphics[width=1\linewidth]{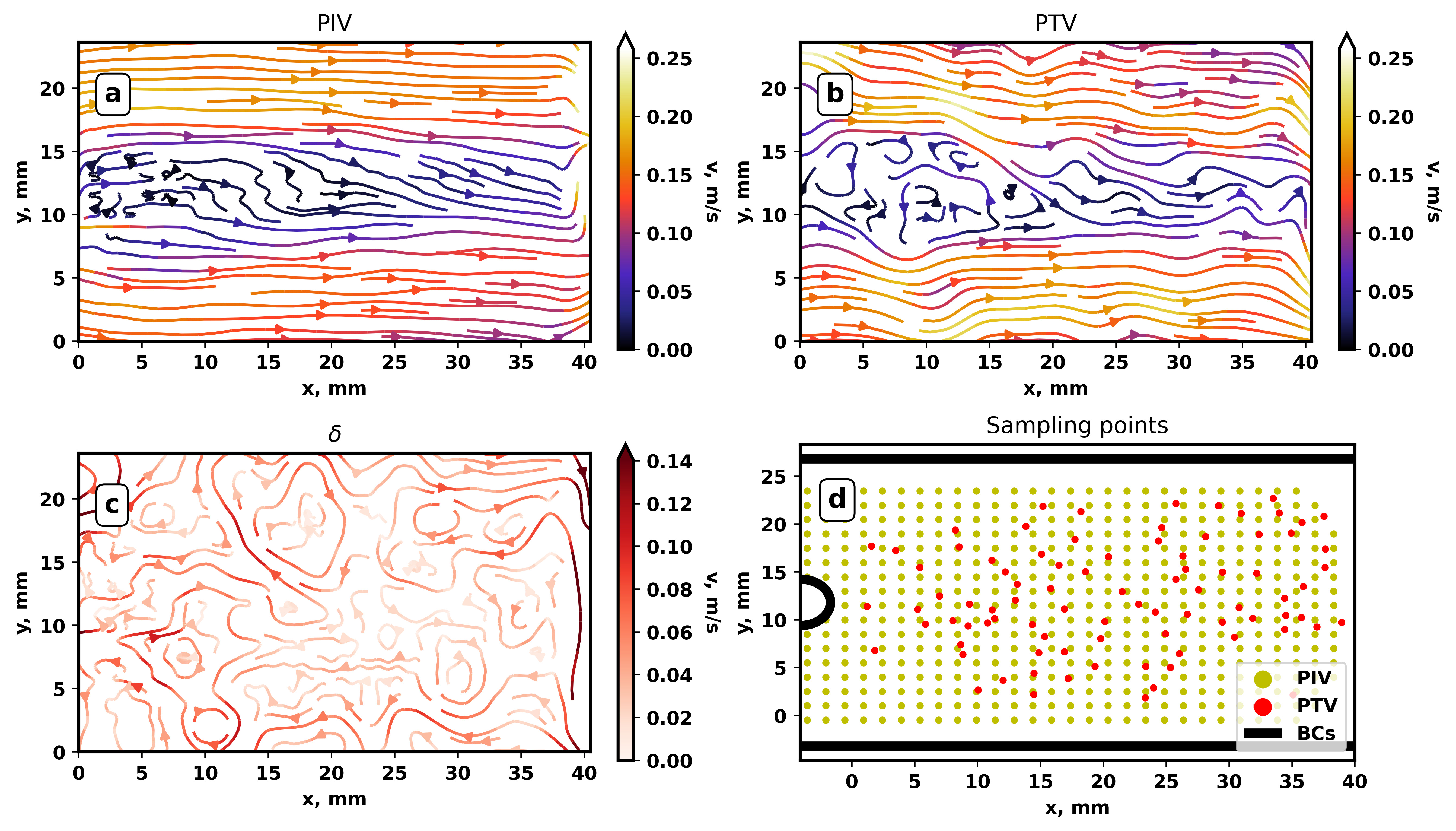}
    \caption{PTV-based reconstruction of a continuous velocity field in the FOV using DFI: (a) DFI-PIV field used for motion prediction; (b) PTV-based continuous field yielded by DFI; (c) the difference between PIV- and PTV-based fields; (d) sampling points for PIV and PTV.}
    \label{fig:PIV_vs_PTV}
\end{figure}

Finally, we can use the PTV results from MHT-X to obtain a continuous velocity field, an example of which can be seen in Figure \ref{fig:PIV_vs_PTV}.

$\vec{v}(\vec{r},t_j)$ over the entire FOV for each frame $t_j$ in an image sequence, with a spatial resolution superior to that of PIV. From our tracking results, we can perform reconstruction using DFI. To do this, we first take each trajectory $\vec{r}(t_k),~ k\in[t_1;t_2]$ and generate a time-parametrized third-order B-spline for it. Here we relax the minimum node threshold for trajectories from 20 to 10, since this amount of points over time is more than sufficient for local PTV. Then the first time derivative is taken for trajectory splines at every frame $t_j$ with the condition $t_j \in (t_1;t_2)$. This way a sparse velocity field is generated for every frame. Afterwards obstacle and wall BC points are added as with DFI for PIV and the PTV + BC point field is interpolated with DFI, yielding $\vec{v}(\vec{r},t_j)$.

As seen in Figure \ref{fig:PIV_vs_PTV}, the resulting $\vec{v}(\vec{r},t_j)$ (b) indeed has a greater resolution than PIV (a) and resolves the finer vortices within the wake much better, while still preserving the larger-scale flow features. The difference between the PIV field (which could be considered as the initial guess) and the refined PTV-based DFI field are shown in (c). Note that we do not perform temporal averaging -- (b) shows an unfiltered instantaneous velocity field reconstructed based only on the points in that frame (d).

\section{Conclusions \& outlook}
\label{sec:conclusions}

To summarize, we have shown improved analysis methods for particle-laden liquid metal flow imaged with neutron radiography, although our approach is readily extendable to other modes of imaging as well as physical systems. Specifically, we have demonstrated how adding a trajectory re-evaluation step to our MHT-X object tracking code, augmenting particle image velocimetry-assisted motion prediction with divergence-free interpolation which enforces boundary conditions and flow incompressibility, as well as upgraded artifact removal methods for raw images results in a significantly improved particle tracking quality.

To prove the point, we assessed neutron images of particle-laden liquid metal flow around a cylindrical obstacle in a flat rectangular channel. From these images, we have been able to derive particle velocity PDFs, velocity oscillation frequency PDFs, trajectory curvature PDFs and spatial maps of trajectory curvature, and also assessed particle residence time for a range of the obstacle Reynolds numbers. We have verified the precision and accuracy of our methods by successfully determining the vortex shedding frequencies for the wake of the obstacle, and replicating the trajectory curvature statistics (i.e. universal algebraic decay of curvature PDFs) expected theoretically, as well as from numerical simulations and experiments. The trends seen in the derived particle number density and curvature spatial maps, as well as velocity PDFs for the considered range of the obstacle Reynolds numbers, are as one would expect. We have also verified with simple estimates that the obtained trajectory curvature statistics should not be too strongly affected by particle collisions, since they are expected to be very rare and rather insignificant in terms of transferred kinetic energy. We have found that the estimated particle collision to initial kinetic energy ratio PDFs also exhibit universal algebraic growth and two distinct regions with different exponents, which is qualitatively similar to what is seen for the curvature PDFs. It was also shown that due to the improved particle tracking velocimetry quality and with divergence-free interpolation which incorporates flow boundary conditions, it is possible to reconstruct instantaneous continuous velocity fields that have a much greater spatial resolution than particle image velocimetry output.

All of the above proves that the methods developed here and in \cite{mht-x-og, birjukovs-particle-EXIF} are safely and readily applicable to investigations of turbulence in liquid metal flow using particles as flow tracers. Speaking of our own future research, we plan to apply what was shown here to new model experiments where a stationary cylindrical obstacle will be replaced with one that can move laterally as it reacts to metal flow, emulating bubble zigzagging motion. Particle tracking will be used to characterize the wake of this bubble model in a thicker channel to enable three-dimensional wake development. Experiments will be performed with different configurations of applied magnetic field.

Despite the already promising results, it is critical to point out avenues for further improvement:

\begin{itemize}[noitemsep,topsep=7.5pt,leftmargin=1.5cm]
    \item Since PIV-assisted motion prediction is one of the key aspects here, PIV quality significantly affects tracking performance. We plan to substitute the current standard PIV approach with a hybrid PIV-optical flow method proposed in \cite{tianshu-piv-optical-flow} (open-source code available) which should provide better results for PIV.
    \item Kalman filter is a prospective approach for improved particle motion prediction that we plan to use as a replacement for spline-based trajectory extrapolation, since it can combine three contributions: PIV- and PTV-predicted motion, as well as predictions derived by solving effective equations of motion for particles (Lagrangian approach).
    \item A multi-level approach for divergence-free interpolation \cite{multi_div_free} seems prospective for further improvement of PIV-based predictions, as well as for reducing the computational complexity thereof -- however, it will have to be adapted to sparse grids first, likely using data thinning methods.
    \item Since particle Reynolds numbers are expected to be low in cases similar to ours \cite{birjukovs-particle-EXIF}, one can use a linear drag model for effective equations of motion for particles; promising methods for solving these equations accurately have been proposed in \cite{div-free-inter-and-motion-prediction} where phase space volume contractivity preservation via splitting methods is used.
    \item For particle-laden flow, it would make sense to implement object cluster detection and joint tracking for improved tracing accuracy when dealing with swarms of coherently moving objects.
\end{itemize}

All of the tools utilized in this paper are open-source. The image processing code is available at \textit{GitHub}: \href{https://github.com/Mihails-Birjukovs/Low_C-SNR_Particle_Detection}{Mihails-Birjukovs/Low\_C-SNR\_Particle\_Detection}. MHT-X can be found at \textit{GitHub} as well: \href{https://github.com/Peteris-Zvejnieks/MHT-X}{Peteris-Zvejnieks/MHT-X}.
There is also a \textit{GitHub} repository for our implementation of divergence-free interpolation \href{https://github.com/Peteris-Zvejnieks/DivergenceFreeInterpolation}{Peteris-Zvejnieks/DivergenceFreeInterpolation}, which is available as a \textit{Python} \href{https://pypi.org/project/Divergence-Free-Interpolant/}{PyPi} package as well.

\section*{Acknowledgements}

This research is a part of the ERDF project ”Development of numerical modelling approaches to study complex multiphysical interactions in electromagnetic liquid metal technologies” (No. 1.1.1.1/18/A/108) and is based on experiments performed at the Swiss spallation neutron source SINQ, Paul Scherrer Institute, Villigen, Switzerland. The authors acknowledge the support from Paul Scherrer Institut (PSI) and Helmholtz-Zentrum Dresden-Rossendorf (HZDR), and specifically express gratitude to Knud Thomsen and Martins Sarma. The work is also supported by a DAAD Short-Term Grant (2021, 57552336) and the ANR-DFG project FLOTINC (ANR-15-CE08-0040, EC 217/3).

\printbibliography[title={References}]

\end{document}